\documentclass[pre,twocolumn,aps,floats]{revtex4}
\usepackage{graphicx,epstopdf}

\newcommand{\csch}{\mbox{csch}}
\newcommand{\sech}{\mbox{sech}}
\newcommand{\re}{\mbox{Re\,}}
\newcommand{\Du}{\mbox{D\hspace{-0.5pt}u}}
\newcommand{\del}{\mbox{\boldmath{$\nabla$}}}
\newcommand{\nb}{\mbox{\bf n}}
\newcommand{\Fb}{\mbox{\bf F}}

\newcommand{\rb}{\mbox{\bf r}}

\begin{document}
%\draft

%\title{Intermediate time scale in the relaxation of
%a dilute electrochemical cell}

\title{ Diffuse-Charge Dynamics in Electrochemical Systems }

\author{Martin Z. Bazant$^{1,2}$, Katsuyo
Thornton$^3$, and  Armand Ajdari$^2$}

\affiliation{
$^1$ Department of Mathematics, Massachusetts Institute of Technology,
Cambridge, MA 02139 \\
$^2$ Laboratoire de Physico-Chimie Th{\'e}orique, UMR ESPCI-CNRS 7083,
10 rue Vauquelin, F-75005 Paris, France\\
$^3$ Department of Materials Science and
Engineering, Northwestern University, Evanston, IL 60201}

% first calculations: June 1999, Bordeaux.
% Began paper: Paris, July 2000.
% First draft without much numerics: Jan 2003
% First COMPLETE draft: Jan 2004.

\date{\today}

\begin{abstract}

The response of a model micro-electrochemical system to a
time-dependent applied voltage is analyzed. The article begins with a
fresh historical review including electrochemistry, colloidal science,
and microfluidics. The model problem consists of a symmetric binary
electrolyte between parallel-plate, blocking electrodes which suddenly
apply a voltage.  Compact Stern layers on the electrodes are also
taken into account. The Nernst-Planck-Poisson equations are first
linearized and solved by Laplace transforms for small voltages, and
numerical solutions are obtained for large voltages.  The ``weakly
nonlinear'' limit of thin double layers is then analyzed by matched
asymptotic expansions in the small parameter $\epsilon = \lambda_D/L$,
where $\lambda_D$ is the screening length and $L$ the electrode
separation. At leading order, the system initially behaves like an RC
circuit with a response time of $\lambda_D L / D$ (not
$\lambda_D^2/D$), where $D$ is the ionic diffusivity, but nonlinearity
violates this common picture and introduce multiple time scales. The
charging process slows down, and neutral-salt adsorption by the
diffuse part of the double layer couples to bulk diffusion at the time
scale, $L^2/D$. In the ``strongly nonlinear'' regime (controlled by a
dimensionless parameter resembling the Dukhin number), this effect
produces bulk concentration gradients, and, at very large voltages,
transient space charge. The article concludes with an overview of more
general situations involving surface conduction, multi-component
electrolytes, and Faradaic processes.

\end{abstract}

\maketitle

%%%%%%%%%%%%%%%%%%%%%%%%%
\section{Introduction }
%%%%%%%%%%%%%%%%%%%%%%%%%

There is rapidly growing interest in micro-electrochemical or
biological systems subject to time-dependent applied voltages or
currents. For example, AC voltages applied at microelectrodes can be
used to pump liquid
electrolytes~\cite{ramos98,ramos99,green00a,gonzalez00,green02,ajdari00,brown01,studer02,mpholo03,ramos03,nadal02b},
to separate or self-assemble colloidal
particles~\cite{yeh97,trau97,faure98,green00b,nadal02a,marquet02,ristenpart03}, and
to manipulate biological cells and
vesicles~\cite{helfrich74,mitov93,pethig96}. Conversely, oscillating
pressure-driven flows can be used to produce frequency-dependent
streaming potentials to probe the structure of porous
media~\cite{journiaux97,reppert01,reppert02}.

A common feature of these diverse phenomena is the dynamics of diffuse
charge in microscopic systems.  Although the macroscopic theory of
neutral electrolytes with quasi-equilibrium double layers is very well
developed in electrochemistry~\cite{bard,newman} and colloidal
science~\cite{hunter,russel,lyklema}, microscopic double-layer
charging at subdiffusive time scales is not as well
understood. Although much progress has been made in various disjoint
communities, it is not so widely appreciated, and some open questions
remain, especially regarding nonlinear effects. The goals of this
paper are, therefore, (i) to review the relevant literature and (ii)
to analyze a basic model problem in considerable depth, highlighting
some new results and directions for further research.

\begin{figure}
\includegraphics[width=3in]{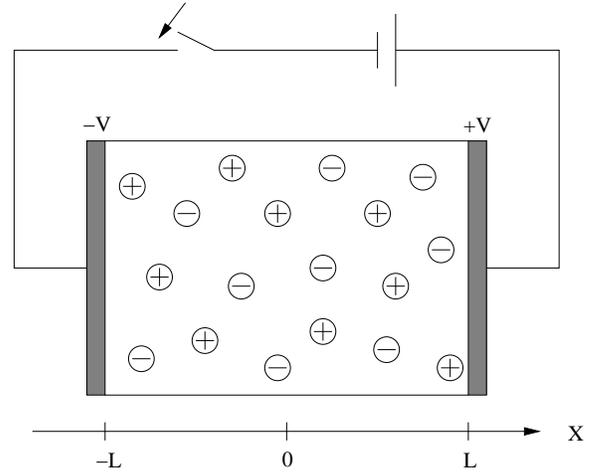}
\caption{ Sketch of the model problem. A voltage $2V$ is suddenly
applied to a dilute, symmetric, binary electrolyte between
parallel-plate, blocking electrodes separated by
$2L$. \label{fig:cartoon} }
\end{figure}

To illustrate the physics of diffuse-charge dynamics, consider the simplest
possible case sketched in Fig.~\ref{fig:cartoon}: a dilute $z$:$z$ electrolyte
suddenly subjected to a DC voltage, $2V$, by parallel-plate blocking
electrodes separated by $2L$. Naively, one might assume a uniform bulk electric
field, $E = V/L$, but the effect of the applied voltage is not so
trivial. Ions migrate in the bulk field and eventually screen it completely
(since ``blocking electrodes'' do not support a Faradaic current).

What is the characteristic time scale of this response? For charge relaxation,
one usually quotes the time, $\tau_D = \lambda_D^2 /D$, for diffusion with a
diffusivity $D$ across one Debye screening length,
\begin{equation}
\lambda_D = \sqrt{\frac{\varepsilon k T}{2z^2e^2C_b}} ,
 \label{eq:lambda}
\end{equation}
where $C_b$ is the average solute concentration, $k$ Boltzmann's
constant, $T$ the temperature, $e$ is the electronic charge, and $\varepsilon$ the permittivity of the
solvent~\cite{hunter,russel,lyklema}. The Debye time, $\tau_D$, is a material
property of the electrolyte, which for aqueous solutions ($\lambda_D
\approx 1-100$ nm, $D \approx 10^3$ $\mu$m$^2$/s) is rather small, in
the range of ns to $\mu$s. More generally, when Faradaic reactions occur (for
a non-blocking electrode), diffuse charge may also vary on the much
slower, geometry-dependent scale for bulk diffusion given by $\tau_L = 
L^2/D$, proportional to the square of the electrode separation.

These two relaxation times, $\tau_D$ for the charge density and
$\tau_L$ for the concentration, are usually presented as the only ones
controlling the evolution of the system, e.g.~ as in the recent
textbooks of Hunter~\cite{hunter} (Ch.\ 8) and Lyklema~\cite{lyklema}
(Chs.\ 4.6c). Dimensional analysis, however, allows for many other time
scales obtained by combining these two, such as the harmonic mean,
\begin{equation}
\tau_c = \sqrt{\tau_D \tau_L} = \frac{\lambda_D L}{D},  \label{eq:tmixed}
\end{equation}
proportional to the electrode separation (not squared). Below, we will
show that this is the primary time scale for diffuse-charge dynamics in
electrochemical cells, although $\tau_D$, $\tau_L$, and other time
scales involving surface properties also play important roles,
especially at large voltages (even without Faradaic processes). The
same applies to highly polarizable or conducting colloidal particles,
where $L$ is the particle size.
%Does a need to be introduced here? (Why not "where $L$ is given by the particle size"?) 

Although the basic charging time, $\tau_c$, is familiar in several scientific
communities~\cite{macdonald70,korn81,dukhin80,dukhin93}, it
is not as widely known as it should be. Recently, it
has been rediscovered as the (inverse) frequency of ``AC pumping'' at
patterned-surface micro-electrodes~\cite{ramos98,ajdari00}. As in the past, its
theoretical justification has sparked some controversy~\cite{scott01,ramos01}
related to the applicability of classical circuit
models~\cite{macdonald90,geddes97} in which $\tau_c$ arises as the ``RC time''
of a bulk resistor in series with a double-layer capacitor (see below).

Here, we attempt to unify and modestly extend a large body of prior
work on diffuse-charge dynamics in the context of our model problem,
paying special attention to effects which undermine the classical
circuit approximation. Going beyond most previous mathematical
studies, we allow for compact-layer capacitance, bulk concentration
polarization, and large voltages outside the linear regime. For the
nonlinear analysis, the method of matched asymptotic
expansions~\cite{bender,hinch,kevorkian} must be adapted for multiple time
scales at different orders of the expansion, so the problem also
presents an opportunity for mathematicians to develop time-dependent
boundary-layer theory.

We begin in section~\ref{sec:history} by reviewing some of the
relevant literature on electrochemical relaxation. In
section~\ref{sec:setup} we state the mathematical problem for a
suddenly applied DC voltage, and in section~\ref{sec:lin} we analyze
the linear response  using Laplace transforms. In
section~\ref{sec:nondim} we non-dimensionalize the problem and
describe numerical solutions, used to test our analytical
approximations. In section~\ref{sec:nonlin}, we derive uniformly valid
asymptotic expansions in the ``weakly nonlinear'' limit of thin double
layers and discuss the connection with circuit models. Apparently for
the first time (for this problem), in section~\ref{sec:higher} we 
analyze higher-order corrections, and in section~\ref{sec:strong} we
briefly discuss the ``strongly nonlinear'' regime at large voltages,
where the expansions are no longer valid.  In section~\ref{sec:concl},
we conclude by briefly discussing extensions to higher dimensions, general
electrolytes, and Faradaic processes and posing some open questions.

\section{ Historical Review }
\label{sec:history}

\subsection{ Electrical Circuit Models }

In electrochemistry, the most common theoretical approach is to
construct an equivalent electrical circuit, whose parameters are fit
to experimental impedance spectra or pulsed-voltage responses, as
recently reviewed by Macdonald~\cite{macdonald90} and
Geddes~\cite{geddes97}. The basic idea of an equivalent circuit is
apparently due to Kohlrausch~\cite{kohlrausch1873} in 1873, and the
first mathematical theory of Kohlrausch's ``polarization capacitance''
was given by Warburg at the end of the nineteenth
century~\cite{warburg1899,warburg1901}. Warburg argued that AC
electrochemical response is dominated by pure diffusion of the active
species and can be described a bulk resistance in series with a
frequency-dependent capacitance, which combine to form the ``Warburg
impedance''. 

Earlier, Helmholtz~\cite{helmholtz1853,helmholtz1879} had
suggested that the solid-electrolyte interface acts like a thin
capacitor, for which he apparently coined the term, ``double
layer''~\cite{bard}. In 1903 Kr\"uger~\cite{kruger1903} unified
Warburg's bulk impedance with Helmholtz' double-layer capacitor in the
first complete AC circuit model for an electrochemical cell, which
forms the basis for the modern ``Randles circuit''~\cite{randles47}.
In this context, the relaxation time for charging of the double layers
has been known to depend on the electrode separation, via the bulk
resistance, for at least a century.

The study of diffuse charge in the double layer was initiated in the
same year by Gouy~\cite{gouy1903}, who suggested that excess ionic
charge in solution near the electrode could be viewed as a
capacitance, $C_D = \varepsilon/\lambda_D$.  He was also the first to
derive Equation~(\ref{eq:lambda}) for $\lambda_D$ (obviously with
different notation) in his original theory of the diffuse double layer
in equilibrium~\cite{gouy1909,gouy1910}. With the availability of
Einstein's relation~\cite{einstein1905} for the mobility, $\mu =
D/kT$, at that time, the DC bulk resistance (per unit area) could have
been calculated as
\begin{equation}
R_b = \frac{V }{J} = \frac{L E_0}{\sigma_b E_0} = \frac{\lambda_D^2
L}{\varepsilon D}  \label{eq:Rb}
\end{equation}
(for a symmetric binary electrolyte of equal mobilities), where $J$ is
the current density and 
\begin{equation}
\sigma_b = \frac{\varepsilon D}{\lambda_D^2} = \frac{ 2(ze)^2 C_0
  D}{kT}  \label{eq:Kb}
\end{equation}
is the bulk conductivity. Therefore, basic time scale in
Eq.~(\ref{eq:tmixed}) has essentially been contained in circuit models
since roughly 1910 as the relaxation time,
\begin{equation}
\tau_c = R_b C_D = \frac{\lambda_D L}{D} = \frac{C_D L}{\sigma_b} =
\frac{\varepsilon L}{\lambda_D \sigma_b}
\end{equation}
although $\tau_c$ was not stated explicitly
as $\lambda_D L /D$ for perhaps another fifty years
~\cite{macdonald70}.

Today, Gouy's screening length bears the name of Debye, who rederived
it in 1923 as part of his seminal work with
H\"uckel~\cite{debye23a,debye23b} on charge screening in bulk
electrolytes, using an equivalent formalism. Debye and H\"uckel solved
for the spherical screening cloud around an ion, and, due to the low
potentials involved, they linearized the transport equations, allowing
them to handle general electrolytes. When Gouy~\cite{gouy1910}
considered the identical problem of screening near a flat, blocking
electrode more than a decade earlier, he obtained exact solutions to
full nonlinear equations for the equilibrium potential profile in
several cases of binary electrolytes, $z_+/z_- = 1$, $2$, and
$\frac{1}{2}$, where $z_+$ and $z_-$ are the cation and anion charge
numbers, respectively. 

A few years later, Chapman~\cite{chapman1913} independently derived
Gouy's solution for a univalent electrolyte, $z_+ = z_- = 1$, the
special case of ``Gouy-Chapman theory'' for which they are both
primarily remembered today. Chapman also gave a simple form for the
charge-voltage relation of the diffuse-layer capacitor in this case,
which, upon differentiation, yields a simple formula for the nonlinear
differential capacitance of the diffuse layer,
\begin{equation}
C_D(\zeta) = \frac{\varepsilon}{\lambda_D} \,
\cosh\left(\frac{ze\zeta}{kT}\right), \label{eq:C_GC}
\end{equation}
where $\zeta$ is the voltage across the diffuse layer in thermal
equilibrium. (Here, we include the trivial extension to a general
$z$:$z$ electrolyte.)  Combining Eqs.~(\ref{eq:Rb}) and
(\ref{eq:C_GC}), we also obtain the basic relaxation time, $\tau_c$,
in Eq.~(\ref{eq:tmixed}) multiplied by a potential-dependent factor in
the usual case of nonzero equilibrium zeta potential (in the absence
of an applied voltage). This factor may be neglected in the
Debye-H\"uckel limit of small potentials, $\zeta \ll kT/ze$, but it
becomes important at large potentials and generally slows down the
final stages of double-layer charging.

More sophisticated models of the double layer were proposed by many
subsequent authors~\cite{delahay,bockris} and incorporated into AC
circuit models for electrochemical
cells~\cite{bard,sluyters70,parsons90}. Naturally, the original ideas of
Helmholtz and Gouy were eventually combined into a coherent
whole. In 1924, Stern~\cite{stern24} suggested   
decomposing the double layer into a ``compact'' (Helmholtz) part within
a molecular distance of the surface and a ``diffuse'' (Gouy) part
extending into the solution at the scale of the screening
length. Physically, the compact layer is intended to describe ions (at
the outer Helmholtz plane) whose solvation
molecules are in contact with the surface, although specifically
adsorbed ions (themselves in contact with the surface) may also be included~\cite{vorsina39}. Regardless of the precise microscopic
picture, however, Stern introduced the compact layer as an intrinsic surface
capacitance, which cuts off the divergent capacitance of
the diffuse layer, Eq.(\ref{eq:C_GC}), at large zeta potentials.

Using this model of two capacitors in series and neglecting
specific adsorption, Grahame~\cite{grahame47} applied Gouy-Chapman
theory for the diffuse part and inferred the nonlinear differential
capacitance of the compact part from his famous experiments on
electrified liquid-mercury drops. Macdonald~\cite{macdonald54} then
developed a mathematical model for double layers at metal electrodes
by viewing the compact layer as a parallel-plate capacitor, as we do
below, although he also allowed its thickness and capacitance to vary
due to electrostriction and dielectric saturation~\cite{grahame50}. 
The reader is referred to
various recent reviews~\cite{macdonald90,geddes97,parsons90}
to
learn how other effects neglected below, such as specific adsorption
and Faradaic processes for non-blocking electrodes, have been included
empirically in modern circuit models.

In spite of a century of research, open questions remain about the
applicability of circuit models~\cite{geddes97}, and even the most
sophisticated fits to experimental data still suffer from
ambiguities~\cite{macdonald90}.  One problem is the somewhat arbitrary
distinction between the diffuse layers and the bulk electrolyte, which
in fact comprise a single, continuous region. Even accepting this
partition, it is clear that the non-uniform evolution of ionic
concentrations in both regions cannot be fully captured by homogeneous
circuit elements~\cite{hollings03}. Another problem is the further
partitioning of the double layer into two (or more) poorly defined
regions at atomic lengths scales, where macroscopic continuum theories
(e.g. for dielectric response) are of questionable
validity~\cite{cooper78}.

\subsection{ Microscopic Transport Models }
\label{sec:hist2}

An alternative theoretical approach, pursued below, is to solve the
time-dependent Nernst-Planck
equations~\cite{nernst1888,nernst1889,planck1890} for ionic transport
across the entire cell (outside any molecular-scale compact layers)
without distinguishing between the diffuse-charge layers and the
quasi-neutral bulk.  Because this ``phenomenological''~\cite{korn81}
approach requires solving Poisson's equation for the mean-field
electrostatic potential (self-consistently generated by the continuum
charge density) down to microscopic (and sometimes atomic) length
scales, it lacks the thermodynamic justification of traditional
macroscopic theories based on bulk electroneutrality and
electrochemical potentials~\cite{newman}. Nevertheless, it addresses
time-dependent charge-relaxation phenomena, which do occur in real
systems, with fewer {\it ad hoc} assumptions than circuit models and
thus may be considered closer to first principles. The use of the
Nernst-Planck equations at scales smaller than the screening length
(but still larger than atomic dimensions) is also supported by the
success of Gouy-Chapman theory in predicting the diffuse-layer
capacitance in a number of experimental systems
(e.g. Refs.~\cite{grahame47,macdonald54}), because the theory is based
on the steady-state Nernst-Planck equations for thermal equilibrium.
The main difficulty in working with the Nernst-Planck equations, aside
from mathematical complexity, is perhaps in formulating appropriate
boundary conditions at the electrode surface, just outside any compact
layers.

Although the response to a suddenly applied DC voltage has been
considered by a few authors in the
linear~\cite{lemay53,buck69,korn77b} and
nonlinear~\cite{korn81,korn78} regimes, as we also do below, much more
analysis has been reported for the case of weak AC forcing, where the
equations are linearized and the time dependence is assumed to be
sinusoidal. These simplifications are made mainly for analytical
convenience, although they have direct relevance for the
interpretation of impedance spectra. An early analysis of this type
was due to Ferry~\cite{ferry48}, who considered the response of a
semi-infinite electrolyte to an oscillating charge density applied at a
electrode surface. Ferry's treatment is formally equivalent to the
classical theory of dielectric dispersion in bulk
electrolytes~\cite{debye28,falkenhagen34}.  Naturally, in both cases
the same time scale, $\tau_D = \lambda_D^2/D$, arises, and the
relaxation of the double layer has no dependence on the macroscopic
geometry.

Ferry's analysis of a single electrode is consistent with the common
intuition that double-layer charging should be a purely microscopic
process, but one might wonder how the electrode could draw charge
``from infinity'' when an infinite electrolyte has infinite
resistance. Indeed, as emphasized by Buck~\cite{buck69} and
Macdonald~\cite{macdonald70} and confirmed by detailed comparisons
with experimental impedance spectra, Ferry's analysis is fundamentally
flawed, starting from the boundary conditions: It is not possible to
control the microscopic charge density at an electrode surface and
neglect its coupling to bulk transport processes; instead, one imposes
a voltage relative to another electrode and observes the resulting
current (or vice versa), while the surface charge density evolves
self-consistently.  

Buck~\cite{buck69} eventually corrected Ferry's analysis to account
for the missing ``IR drop'' across two electrodes which imposes the
initial surface charge density (or, alternatively, voltage). Nevertheless, the
physical picture of a double-layer responding locally to ``charge
injection'', independent of bulk transport processes, persists to the
present day.  For example, recent textbooks on colloidal science
(Hunter~\cite{hunter}, p.\ 408; Lyklema~\cite{lyklema}, p.\ 4.78)
present a slightly different version of Ferry's analysis (attributed
to O'Brien) as the canonical problem of ``double-layer relaxation'':
the response of a semi-infinite electrolyte to a suddenly imposed,
constant surface charge density. This gives some insight into high
frequency dielectric dispersion of non-polarizable colloids (the usual
case), but it is not relevant for polarizable particles and
electrodes. Three decades after Buck and Macdonald, it is worth
re-emphasizing the fundamental coupling of double-layer charging of
bulk transport in finite, polarizable systems.

The mathematical theory of AC response for a finite, two-electrode
system began with Jaff\'e's analysis for
semiconductors~\cite{jaffe33,jaffe52} and was extended to liquid
electrolytes by Chang and Jaff\'e~\cite{chang52}. A number of
restrictive assumptions in these studies, such as a uniform electric
field, were relaxed by Macdonald~\cite{macdonald53} for semiconductors
and electrolytes and independently by Friauf~\cite{friauf54} for ionic
crystals. These authors, who gave perhaps the first complete
mathematical solutions, also allowed for bulk generation/recombination
reactions, which are crucial for electrons and holes in
semiconductors. Subsequent authors mostly neglected bulk reactions in
studies of
liquid~\cite{macdonald70,buck69,baker68,macdonald74a,macdonald74b} and
solid~\cite{beaumont67,korn77b} electrolytes, while focusing on other
effects, such as arbitrary ionic valences and the compact layer.

Although it is implicit in earlier work, Macdonald~\cite{macdonald70}
first clearly identified the geometry-dependent time scale, $\tau_c =
\lambda_D L/D$, (in this form) as governing the relaxation of an
electrochemical cell. It was also derived independently by Kornyshev
and Vorontyntsev~\cite{korn77b,korn78} in the Russian literature on
solid electrolytes with one mobile ionic species~\cite{korn81}. With
Itskovich~\cite{korn77a}, these authors also modeled the compact-layer
capacitance via a mixed Dirichlet-Neumann condition on the
Nernst-Planck equations. This classical boundary
condition~\cite{bard}, also used below, introduces another length
scale, $\lambda_S$, the effective width of the Stern layer, which also
affects the time scales for electrochemical relaxation.

Other important surface properties have also been included in
mathematical analyses of AC response. For example, several recent
studies of blocking electrodes have included the effect of a nonzero
equilibrium zeta potential (away from the point of zero
charge)~\cite{hollings03,gunning95,scott00a,scott00b}, building on the
work of Delacey and White~\cite{delacey82}. A greater complication is
to include Faradaic processes at non-blocking electrodes through
boundary conditions of the Butler-Volmer
type~\cite{newman,delahay,bockris}, as suggested by
Levich~\cite{levich49} and Frumkin~\cite{frumkin55}. This approach has
been followed in various analyses of AC response around base states of
zero~\cite{korn81,korn77a,korn78,macdonald76a,macdonald76b,macdonald77}
and nonzero~\cite{bonnefont01,bonnefont_thesis} steady Faradaic
current. Numerical solutions of the time-dependent Nernst-Planck
equations have also been developed for AC response and more general
situations~\cite{delacey82,brumleve78,murphy92}, following the
original work of Cohen and Cooley~\cite{cohen65}.
% The effect of surface
% roughness has also been studied recently~\cite{pajkossy94}, although still
% relatively little mathematical analysis of electrochemical relaxation in two or
% three dimensions has been done.

\subsection{ Colloids and Microfluidic Systems }
\label{sec:colloids}

Diffuse-charge dynamics occurs not only near electrodes, but also
around colloidal particles and in microfluidic systems, where the
coupling with fluid flow results in time-dependent, nonlinear
electrokinetic phenomena. This review may be the first to unify some
of the fairly disjoint literatures on diffuse-charge dynamics in these
areas with the older literature in electrochemistry discussed
above. Compared to the latter, more sophisticated mathematical
analyses are often done in colloidal science and
electro-microfluidics, starting from the Nernst-Planck equations for
ion transport and the Navier-Stokes equations for fluid mechanics in
two or three dimensions. On the other hand, with the notable exception
of the Ukrainian school~\cite{dukhin93,lyklema,dukhin02,murtsovkin96}, less
attention is paid to surface properties, and simple boundary
conditions are usually assumed (constant zeta potential and complete
blocking of ions) which exclude diffuse-charge dynamics.

This might explain why the material time scale, $\tau_D$, is
emphasized as the primary one for double-layer relaxation around
colloidal particles~\cite{hunter,russel,lyklema}, although the mixed
time scale, $\tau_c = (L/\lambda_D)\tau_D$, has come to be recognized
as controlling bulk-field screening by
electrodes~\cite{hollings03,gunning95,scott00a,scott00b}. This
thinking can be traced back to the seminal work of Debye and
Falkenhagen~\cite{debye28,falkenhagen34} on dielectric dispersion in
bulk electrolytes, mentioned above. In that context, when a background
field $E_b$ is applied to an electrolyte, the relevant geometrical
length is the size of the screening cloud around an ion, $L =
\lambda_D$, over which a voltage, $E_b \lambda_D$, is effectively
applied. The relevant RC time for the polarization of the
screening cloud is then, $\tau_c = \lambda_D \lambda_D/D =
\tau_D$. The possible role of geometry is masked by the presence of
only one relevant length scale, $\lambda_D$.

For colloidal particles, which are usually much larger than the
double-layer thickness, the second time scale, $\tau_a = a^2/D$, for
bulk diffusion around a particle of radius, $a$, becomes important,
especially in strong fields. If there is significant surface conduction
or the particle is conducting, the ``RC'' time scale, $\tau_c =
\lambda_D a/D$, can also become important. In general, double-layer
relaxation is thus sensitive to the size and shape of the
particle. Although it is largely unknown (and rarely cited) in the
West, many effects involving non-uniform double-layer polarization
around colloidal particles have been studied 
under the name, ``non-equilibrium electric surface
phenomena''~\cite{dukhin80}, as recently reviewed by S.\ S.\
Dukhin~\cite{dukhin93,dukhin02}.

The colloidal analog of our model problem involving a blocking
electrochemical cell is that of an ideally polarizable, metal particle
in a suddenly applied background electric field. This situation has
received much less attention than the usual case of non-conducting
particles of fixed surface charge density, but it has an interesting
history.  The non-uniform polarization of the double layer for a metal
particle was perhaps first described by Levich~\cite{levich}, using
Helmholtz' capacitor model.  Simonov and
Shilov~\cite{simonov73a,simonov77} later considered diffuse charge and
showed that the metal particle acquires an induced dipole moment
opposite to the field over the time scale, $\tau_c = \lambda_D a /D$,
as bulk conduction transfers charge from the part of the double-layer
facing away from the field to the part facing toward the field. The
two hemispheres may be viewed as capacitors coupled through a
continuous bulk resistor~\cite{simonov77}, as in the RC circuit model
of DC electrochemical cells described above. The charging process
continues until the redistribution of diffuse-charge completely
eliminates the normal component of the electric field, responsible for
charging the double layer.

Diffuse-charge dynamics is important in the context of colloids
because it affects electrokinetic phenomena. In the metal-sphere
example, the remaining tangential component of the field interacts
with the non-uniform induced diffuse charge (and zeta potential) to
cause nonlinear electro-osmotic flows~\cite{murtsovkin96,gamayunov86},
which causes hydrodynamic interactions between colloidal
particles. Although these flows have little effect on the
electrophoresis of charged polarizable particles in uniform DC
fields~\cite{simonov73b,simonova76b}, they significantly affect
dielectrophoresis in nonuniform AC fields~\cite{shilov81,simonova01},
where the time-dependence of double-layer relaxation also plays an
important role. 

These developments followed from pioneering studies of S.\ S.\ Dukhin, B.\ V.\ Deryagin, and collaborators~\cite{dukhin80,dukhin93,deryagin80,dukhin74} 
on the effects of surface conduction and concentration
gradients on electrical polarization and electro-osmotic flows around
highly charged non-conducting particles, which was also extended to
polarizable particles~\cite{murtsovkin96}. (Similar ideas were also
pursued later in the Western literature, with some new
results~\cite{hunter,obrien83,hinch84,prieve84,anderson89}.)  Earlier still,
Bikerman~\cite{bikerman33,bikerman35,bikerman40} presented the
original theory of surface conduction in the double layer, and
Overbeek~\cite{overbeek43} first calculated in detail the effect
of non-equilibrium double-layer polarization on electrophoresis.
%  Some
% of these exotic NESP involving diffuse-charge dynamics may have
% relevance for recent experiments on colloidal self-assembly near
% electrodes in AC fields~\cite{yeh97,trau97}.

Diffuse-charge dynamics has begun to be exploited in microfluidic
devices, albeit without the benefit of the prior literature in
electrochemistry and colloidal science discussed above. In a series of
recent papers, Ramos and collaborators have predicted and observed
``AC electro-osmosis'' at a pair of blocking
micro-electrodes~\cite{ramos98,ramos99,green00a,gonzalez00,green02}. Their
simple explanation of double-layer dynamics~\cite{ramos98,ramos99},
supported by a mathematical analysis of AC response in two
dimensions~\cite{gonzalez00}, is similar to that of Simonov and Shilov
for a metal particle in an AC field~\cite{simonov77}, and the
resulting electro-osmotic flow is of the type described by Gamanov et
al.\ for metal particles~\cite{gamayunov86}. An important difference,
however, is that AC electro-osmosis occurs at fixed micro-electrodes,
whose potentials are controlled externally, as opposed to free
colloidal particles.  Ajdari~\cite{ajdari00} has proposed a similar
means of pumping liquids using AC voltages applied at an array of
micro-electrodes, where broken symmetries in surface geometry or
chemistry generally lead to net pumping past the array, as observed in
subsequent experiments~\cite{brown01,studer02,mpholo03,ramos03}. These
are all examples of the general principle of ``induced-charge
electro-osmosis''~\cite{bazant03,squires03}, where diffuse-charge
dynamics at polarizable surfaces (not necessarily electrodes) is used
to drive micro-flows with AC or DC forcing. Clearly, the full range of
possible microfluidic applications of time-dependent nonlinear
electrokinetics has yet to be explored.

\subsection{ The Limit of Thin Double Layers}

All of the analytical studies cited above that go beyond linear
response (and most that do not) are based on the thin-double-layer
approximation, $\lambda_D \ll L$. In this limit, the bulk electrolyte
remains quasi-neutral, and the double layer remains in thermal
quasi-equilibrium, even with time dependent forcing (slower than
$\tau_D$)~\cite{newman,lyklema,hunter,russel,dukhin93}. The same limit
also justifies the general notion of circuit models for the diffuse
part of the double layer and, in the absence of concentration
gradients, the neutral bulk region.

As first shown by Grafov and Chernenko~\cite{grafov62,chernenko63} in
the Soviet Union and by Newman~\cite{newman66} and
Macgillivray~\cite{macgillivray68} in the United States, the thin
double-layer approximation for electrochemical cells can be given
``firm'' (but not necessarily ``rigorous'') mathematical justification
by the method of matched asymptotic
expansions~\cite{bender,hinch,kevorkian} in the small parameter,
$\epsilon = \lambda_D/L$. For steady Faradaic conduction, the usual
leading-order approximation involves a neutral bulk with charged
boundary layers of $O(\epsilon)$ dimensionless width, which has since
been established rigorously in a number of studies by
mathematicians~\cite{rubinstein90,henry89,louro91,henry95,park97,barcilon97}.  The
standard asymptotic approximation breaks down, however, near Nernst's
diffusion-limited current, where the concentration at the cathode
vanishes. At the limiting current~\cite{smyrl67}, the boundary layer
expands to $O(\epsilon^{2/3})$ width, while at still larger
currents~\cite{rubinstein79}, a layer of ``space charge'' extends out
to $O(1)$ distances into the bulk region, although the effect of realistic
boundary conditions (Faradaic processes, compact layer, etc.) remains
to be studied in these exotic regimes. Matched asymptotic expansions
are also beginning to be used for time-dependent electrochemical
problems with Faradaic processes~\cite{bonnefont01,bonnefont_thesis}
below the limiting current.

Perhaps because it originated in fluid mechanics~\cite{kevorkian}, the
method of matched asymptotic expansions has been used extensively
in colloidal science and microfluidics
~\cite{gonzalez00,dukhin93,dukhin74,obrien83,hinch84,prieve84,anderson89,dukhin69,dukhin70},
albeit with varying degrees of mathematical rigor. In any case, the
advantages of the technique are (i) to justify the assumption 
of equilibrium
structure for the double layers (at leading order), regardless of
transport processes in the neutral bulk, and (ii) to view the double
layers as infinitely thin at the bulk length scale, which is
particularly useful in multidimensional problems. For statics or
dynamics at the bulk diffusion time, it is usually possible to
construct uniformly valid approximations by adding the inner and
outer solutions and subtracting the overlap.

The thin double layer approximation is ``asymptotic'' as $\epsilon
\rightarrow 0$, which means that the ratio of the approximation to the
exact solution approaches unity for sufficiently small $\epsilon$,
with all other parameters held fixed. For any fixed $\epsilon > 0$ (no
matter how small), however, the approximation breaks down at
sufficiently large voltages. The general
criterion
\begin{equation}
\frac{\lambda_D}{a} \, \cosh\left(\frac{ze\zeta}{2kT}\right)\ll 1
\label{eq:valid}
\end{equation}
is often quoted for the validity of Smoluchowski's formula for the
electrophoretic mobility of a thin-double-layer
particle~\cite{hunter}, as justified by numerical
calculations~\cite{obrien78}. This is related to S. S. Duhkin's
seminal work on double-layer distortion around a spherical
particle~\cite{dukhin93,dukhin74,dukhin70}: In the case of highly charged
particles, $\zeta \gg kT/ze$, the ``Dukhin
number'' $\Du$ (which he called ``$Rel$'') controls corrections to
the thin-double-layer limit, $\Du=0$.

The Dukhin number is defined as the ratio of the double-layer surface
conductivity, $\sigma_s$, to the bulk conductivity, $\sigma_b$, in
Eq.~(\ref{eq:Kb}) per geometrical length, $a$: $\Du =
\sigma_s/\sigma_b a$.  Although its effect on electrophoresis was
first explored in detail by Dukhin, the same dimensionless group was
defined a few decades earlier by Bikerman~\cite{bikerman40}, who also
realized that it would play a fundamental role in electrokinetic
phenomena.  In a symmetric binary electrolyte with equal
diffusivities, the Dukhin number can be put in the simple
form,
\begin{eqnarray}
\Du &=& \frac{2\lambda_D(1+m)}{a} \left[ \cosh\left(
\frac{ze\zeta}{2kT}\right) - 1 \right] \nonumber \\
&  =&  \frac{4\lambda_D(1+m)}{a} \sinh^2\left(
\frac{ze\zeta}{4kT}\right).   \label{eq:Du}
\end{eqnarray}
where 
\begin{equation}
m = \left(\frac{kT}{ze}\right)^2 \frac{2\varepsilon}{\eta D} 
\end{equation}
is a dimensionless number giving the relative importance of
electro-osmosis compared with electro-migration and diffusion in surface
conduction, and $\eta$ is the viscosity. This form is due to Deryagin
and Dukhin~\cite{deryagin69}, who generalized Bikerman's original
results~\cite{bikerman33,bikerman35} to account for electro-osmotic
surface conductance ($m>0$). For $\Du \ll 1$ the
double-layer remains in its equilibrium state at constant zeta
potential, but for $\Du \gg 1$ it becomes distorted as surface
conduction draws current lines into the double layer.  For a detailed   
pedagogical discussion, we refer to Lyklema~\cite{lyklema}.

It is interesting to note that (at least at large zeta potentials) the
Dukhin number is similar to the ratio of the effective RC
time, $\tau_c(\zeta)$, away from the point of zero charge ($\zeta \neq
0$) to the bulk diffusion time, $\tau_a$:
\begin{equation}
\frac{\tau_c(\zeta)}{\tau_a} = \frac{\lambda_D}{a} \,
\cosh\left(\frac{ze \zeta}{kT}\right)
\end{equation}
where we have used Eqs.~(\ref{eq:Rb})--(\ref{eq:C_GC}).  Moreover, the
usual condition (\ref{eq:valid}) for the validity of the thin
double-layer approximation in quasi-steady electrokinetic problems is
also a statement about time scales: $\tau_c(\zeta) \ll \tau_L$. When
this condition is violated, the usual RC charging dynamics is
slowed down so much by nonlinearity that bulk diffusion may complicate
the picture. Whether this does in fact occur depends on if the
nonlinearity is strong enough to cause significant concentration
depletion in the bulk for a given geometry and forcing. Understanding
this issue requires going beyond leading order in asymptotic analysis,
which is not trivial.

In spite of extensive work on the asymptotic theory of diffuse-charge
dynamics, difficult open questions remain.  The leading-order
thin-double-layer approximation is well understood in many cases, but
higher-order corrections have been calculated in only a few heroic
instances, such as the asymptotic analysis of diffusiophoresis by
Prieve et al.~\cite{prieve84}. Moreover, such detailed analysis has
mostly (if not exclusively) been done for quasi-steady problems. For
time-dependent problems of double-layer charging, it seems that
higher-order terms in uniformly-valid matched asymptotic expansions
have never been calculated.  

Even the leading-order behavior is poorly understood when the {\it
induced} zeta potential is large enough to violate the condition
(\ref{eq:valid}). In that case, the effective Dukhin number varies
with time, as the total zeta potential evolves in time and space. On
the other hand, the Russian literature on non-equilibrium
electro-surface phenomena at large $\Du$ mostly pertains to highly
charged particles in weak fields, where the constant {\it equilibrium}
zeta potential is large, but the time-dependent induced zeta potential
is small.

Below, we begin to explore these issues in the much simpler context of
a one-dimensional problem involving parallel-plate electrodes, which
excludes surface conduction and electro-osmotic flow.  We shall see
that this requires extending standard boundary-layer theory, which
deals with multiple length scales, to account for simultaneous
multiple time scales.  Before examining the nonlinear theory,
however, we state the mathematical model and study its exact solution
in the linear limit of small potentials.

%%%%%%%%%%%%%%%%%%%%%%%%%%%%%%%%%%%%
\section{ The Basic Mathematical Problem }

\label{sec:setup}
%%%%%%%%%%%%%%%%%%%%%%%%%%%%%%%%%%%%%%%%%

As the simplest problem retaining the essential features of
diffuse-charge dynamics, we consider a dilute, completely dissociated
$z\!:\!z$ electrolyte, limited by two parallel, planar, blocking
electrodes at $X=\pm L$. We describe the concentrations of the charged
ions by continuum fields, $C_{\pm}(X,\tau)$, which satisfy the
Nernst-Planck equations,
\begin{equation}
\frac{\partial C_\pm}{\partial \tau} = - \frac{\partial}{\partial X} \left[
-D \frac{\partial C_\pm}{\partial X} \mp \mu ze C_\pm \frac{\partial
  \Phi}{\partial X}
\right]   \label{eq:eqdim}
\end{equation}
(without generation/recombination reactions), where $\Phi$ is the
electrostatic potential, which describes the Coulomb interaction in a
mean-field approximation.  For simplicity we assume that the diffusion
coefficients of the two ionic species are equal to the same constant,
$D$, and obtain the mobility, $\mu$, from the Einstein relation, $\mu
= D/k T$. The total ionic charge density, $\rho_e$, controls the
spatial variation of the potential, $\Phi$, through Poisson's 
equation,
\begin{equation}
- \varepsilon \frac{\partial^2 \Phi}{\partial X^2} = \rho_e =  ze (C_+ - C_-)
\label{Poi}
\end{equation}
where $\varepsilon$ is the dielectric permittivity of the solvent,
assumed to be a constant.

As described above, we focus on ``ideally polarizable'' or ``completely
blocking'' electrodes without Faradaic processes, so the ionic fluxes have to
vanish there:
\begin{equation}
F_{\pm} = -D \frac{\partial C_{\pm}}{\partial X} \mp \frac{ze D}{k_BT} C_{\pm}
\frac{\partial \Phi}{\partial X}
 =0, \ \ \mbox{for } X=\pm L.
\end{equation}
The Faradaic current density, $J = ze(F_+ - F_-)$, also vanishes at
the electrodes, although it can be nonzero elsewhere as diffuse charge
moves around inside the cell. We also take into account the intrinsic
capacitance of the electrode surface through a mixed boundary
condition for the potential~\cite{bard,korn77a,bonnefont01}. The surface
capacitance may represent a Stern layer of polarized solvent molecules~\cite{stern24}
and/or a dielectric coating on the electrode~\cite{macdonald54}. If $V_{\pm}(t)$ is the external
potential imposed by the external circuit on the electrode at $X=\pm
L$, then we assume
\begin{equation}
\Phi = V_{\pm} \mp \lambda_S \frac{\partial
\Phi}{\partial X} , \ \ \mbox{ at } X = \pm L,
\label{BCpot}
\end{equation}
where $\lambda_S$ is an effective thickness for the compact part of
the double layer. For a simple dielectric layer, this is equal to its
actual thickness times the ratio, $\varepsilon/\varepsilon_S$, of dielectric
constants of the solvent and the Stern layer, $\varepsilon_S$.

In order to study nonlinear effects and avoid imposing a time
scale, we consider the response to a step in voltage (a suddenly applied DC voltage),
rather than the usual case of weak AC forcing. For times $\tau <0$, no
voltage is applied, and we assume no spontaneous charge accumulation at the
electrodes.  The initial ionic
concentrations are uniform, $C_{\pm}(X,\tau <0)=C_0$. For $\tau> 0$, a
voltage difference $2V$ is applied between the two electrodes,
$V_{\pm}(\tau >0)=\pm V$, and we solve for the evolution of the
concentrations and the potential. As $\tau\rightarrow\infty$, the bulk
electric field at the center, $|E(0,\tau)| = \partial \Phi /
\partial X$, decays from its initial value, $V/L$, to zero, due to screening by
diffuse charge which is transferred from the right side of the cell
($0 < X < 1$) to the left ($-1 < X < 0$). The relaxation is complete
when the Faradaic current decays to zero in steady state, from its
initial uniform value, $J(X,0) = J_0 = - \sigma_b V/L= -2 (ze)^2 C_0 D V/ kT L$.

%%%%%%%%%%%%%%%%%%%%%%%%%%%%%%%%
\section{ Linear Dynamics }
\label{sec:lin}
%%%%%%%%%%%%%%%%%%%%%%%%%%%%%%%%%%%%%%%%

\subsection{ Transform Solution for Arbitrary $\lambda_D$, $\lambda_S$,
$L$ }

For applied potentials much smaller than the thermal voltage, $V \ll
k_BT/ze$, the equations can be linearized, $C_{\pm}= C_0 +\delta
C_{\pm}$, so that the ionic charge density,
$\rho_e=ze(C_+-C_-)=ze(\delta C_+ -\delta C_-)$, obeys the
Debye-Falkenhagen equation~\cite{debye28},
\begin{equation}
\frac{1}{D} \frac{ \partial \rho_e}{\partial \tau}
 \approx \frac{\partial^2 \rho_e}{\partial X^2} - \kappa^2 \rho_e
\label{Deb}
\end{equation}
where $\kappa=\lambda_D^{-1}$ is the inverse screening length.
This equation can also be written as a conservation law,
\begin{equation}
\frac{\partial \rho_e}{\partial \tau} = - \frac{\partial J_e}{\partial X}
\end{equation}
in terms of the linearized total ionic electrical current,
\begin{equation}
J_e \approx - D\frac{\partial \rho_e}{\partial X} - D \kappa^2\varepsilon
\frac{\partial \Phi}{\partial X}
\end{equation}
which vanishes at the blocking electrodes, $X=\pm L$.

To solve the model problem, which involves a step-potential in time, it is
convenient to use Laplace transforms, defined by
\begin{equation}
{\hat f}(S) = \int_0^\infty d\tau\,\, e^{-S\tau} f(\tau).
\end{equation}
As $\rho_e(X)=0$ for $\tau<0$, the Laplace transforms of Eqs.~(\ref{Poi}) and
(\ref{Deb}) are
\begin{eqnarray}
 \frac{\partial^2 \hat \rho_e}{\partial X^2} &=&  k^2 {\hat \rho_e}
\label{hDeb} \\
-\varepsilon \frac{\partial^2 \hat\Phi}{\partial X^2} & = & {\hat \rho_e}
\label{hPoi}
\end{eqnarray}
where 
\begin{equation}
k(S)^2= \frac{S}{D} + \kappa^2.
\end{equation}
The general antisymmetric solution to Eq.~(\ref{hDeb}) is,
\begin{equation}
\hat{\rho}_e(X,S)= A \sinh(k X)
\end{equation}
for some constant $A(S)$, which, substituting into Eq.~(\ref{hPoi})
and integrating, yields,
\begin{equation}
- \epsilon_w \frac{\partial \hat{\Phi}}{\partial X}(X,S) =
\frac{A}{k}  \cosh(k X) + B,  \label{ES}
\end{equation}
where the constant $B(S)$, determined by $\hat{J_e}(\pm L,S) = 0$, is given by
\begin{equation}
B = A k \cosh(\kappa_S L) (\kappa^{-2} - k^{-2}) .
\end{equation}
Integrating Eq.~(\ref{ES}) again and enforcing antisymmetry yields the
Laplace transform of the potential,
\begin{equation}
{\hat \Phi}(X,S) = - A \frac{\cosh(k L)}{\varepsilon k^2}
\left( \frac{\sinh(k X)}{\cosh(k L)} + \frac{k S X}{\kappa^2 D} .
\right)
\end{equation}
The remaining constant,
\begin{equation}
A= \frac{-k^2 \varepsilon V S^{-1} \sech(k L) } {
\tanh(kL)+\lambda_S k +\frac{kSL}{\kappa^2 D}
\left(1+\frac{\lambda_S}{L}\right) }
%\right)^{-1}
\end{equation}
is determined by the Stern-layer boundary condition,
Eq.~(\ref{BCpot}).

\subsection{ Long-time Exponential Relaxation }

There is a great deal of information about transients in the Laplace
transform of exact solution to the linear problem. For times much
smaller than the Debye time, $\tau \ll \tau_D = \lambda_D^2/D$ (or $S
\gg \kappa^2 D$), there is no significant response, so we are mainly
interested in the response at longer times, $\tau \gg \tau_D$ (or $S
\ll \kappa^2 D$). There are many ways to see that this is generally an
exponential relaxation dominated by the mixed time scale discussed
above, $\tau_c = \lambda_D L/D$, although several other time scales
allowed by dimensional analysis also play a role.

\subsubsection{ Diffuse Charge Density at a Surface }

Let us focus on one quantity, for example, the Laplace transform of
the charge density at the anode, ${\hat \rho_e}(X=L,S)$.  The exact
formula is
\begin{equation}
{\hat \rho_e}(L,S) = A \sinh(k L) ,
%= \frac{-k^2 \epsilon_w V S^{-1} \tanh(k L) }
%{ \tanh(kL)+\lambda_S k +\frac{kSL}{\kappa^2 D} \left(1+\frac{\lambda_S}{L}\right)
%}.
\end{equation}
which is difficult to invert analytically. (Keep in mind that $k$
depends on $S$.) For times much longer than the Debye time, we
consider the limit, $S \ll \kappa^2 D$, in which the Laplace transform
takes the much simpler asymptotic form,
\begin{equation}
{\hat \rho_e}(L,S) \sim \frac{K_{\rho}\, S^{-1}}{1 + \tau_{\rho} S}
\end{equation}
where
\begin{equation}
K_{\rho} = -\frac{\varepsilon V \kappa^2}{1 + \kappa \lambda_S \coth(\kappa L) }
\end{equation}
and
\begin{equation}
\tau_{\rho} = \frac{L}{\kappa D} \, \left[ \frac{\coth(\kappa L)\left(1 +
\frac{3\lambda_S}{2L}\right) - \frac{1}{2} \kappa \lambda_S \csch^2(\kappa L)
- \frac{1}{\kappa L}}{1 + \kappa \lambda_S \coth(\kappa L) } \right]
\label{eq:taucrho}
\end{equation}
Since the Laplace transform of $1-\exp(-\tau/\tau_o)$ is
$S^{-1}/(1+S\tau_o)$, this result clearly shows that the buildup of
the charged screening layer occurs exponentially over a
characteristic response time given by Eq.~(\ref{eq:taucrho}),
%  In the limit of thin Debye
% layers $L \gg \kappa^{-1}$, this formula simplifies further,
% \begin{equation}
% {\hat \rho_e}(z=L,S\to 0) \simeq -\frac{\kappa^2 \epsilon_w
% V}{(1+\lambda_s\kappa)}\,\, \frac{S^{-1}} { 1+S \frac{L
% (1+\lambda_s/L)}{\kappa D(1+\lambda_s\kappa)}}
% \end{equation}
which is of order, $L/\kappa D = \lambda_D L /D = \tau_c$, for both thin and
thick double layers. Corrections introduce other mixed scales involving the
Stern length, such as $\lambda_S L/D$ and $\lambda_S \lambda_D/D$, as well as
the Debye time, $\lambda_D^2/D$.

Note that the same time scale can also be seen in the linear response to a weak
oscillatory potential, $V_{\pm}=\pm V \re (e^{i\omega \tau})$, which naturally
leads to
\begin{equation}
\rho_e(L,\tau) \sim K_\rho \, \re\left( \frac{e^{i\omega\tau}}{1 +
  i\omega\tau_\rho} \right)
\end{equation}
for frequencies well below the Debye frequency, $\omega \ll \omega_D =
D/\lambda_D^2$. Similar results for AC response near the point of zero charge
have been obtained by many authors, as cited above. The characteristic
frequency, $\omega_c = 1/\tau_c \approx D/\lambda_D L$, also arises the context
of AC electro-osmotic fluid pumping near
micro-electrodes~\cite{ramos98,ajdari00}, because diffuse-layer charging
controls the time-dependence of the effect.

\subsubsection{ Total Diffuse Charge in an Interface }

We now show that the same form of long-time exponential relaxation, with a
somewhat {\it different} characteristic time, also holds for other quantities,
such as the total diffuse charge near the cathode,
\begin{equation}
Q(t) = \int_{-L}^0 \rho(X,t) dX ,
\end{equation}
which plays a central role in the nonlinear analysis below. In the
limit of thin double layers, this is simply the total interfacial
charge (per unit area) of the diffuse part of the double layer. Here
we consider the total diffuse charge near a surface more generally,
even when the Debye screening length is much larger than the electrode
separation. In the latter case, the concept of an ``interface'' is not
well defined, since the two sides of the cell interact very strongly,
but we can still study the overall separation of diffuse charge caused
by the applied voltage.

Using Eqs.~(\ref{hPoi}) and (\ref{ES}), the Laplace transform of the
total cathodic charge is,
\begin{equation}
\hat{Q}(S) = A k^{-1}\left[1 - \cosh(k L)\right] .
\end{equation}
Once again, this is difficult to invert analytically, so we focus on
the long-time limit,
\begin{equation}
\hat{Q}(S) \sim \frac{K_Q S^{-1}}{1 + \tau_Q S} ,
\end{equation}
for $S \ll \kappa^2 D$, where
\begin{equation}
K_Q = \frac{ \varepsilon V \kappa \left[1 - \sech(k L)\right]
}{\tanh(\kappa L)  + \kappa \lambda_S}
\end{equation}
and
\begin{eqnarray}
\tau_Q &=& \frac{L}{\kappa D} \left\{ \frac{1 + \frac{1}{2}\sech^2(\kappa
      L) + \frac{3\lambda_S}{2L}}{\tanh(\kappa L) + \kappa \lambda_S}
       \right. \nonumber \\
& & \left. - \frac{\sech(\kappa L)\tanh(\kappa
      L)}{2\left[1 - \sech(\kappa L)\right]} - \frac{1}{2\kappa L} \right\}   \label{eq:taucQ}
\end{eqnarray}
In the limit of thin double layers, the same basic time scale, $\tau_c
= L/\kappa D = \lambda_D L /D$, arises as in the case of the surface
charge density. A subtle observation is that the relaxation of the
total interfacial charge, although still exponential, has a somewhat
different time scale as a function of $\epsilon = \lambda_D/L$ and
$\delta = \lambda_S/\lambda_D$. (See Fig.~\ref{fig:tc} below.) This
apparently new result shows that charging dynamics has a nontrivial
dependence on time and space, even for very weak potentials.

\begin{figure}
\includegraphics[width=3in]{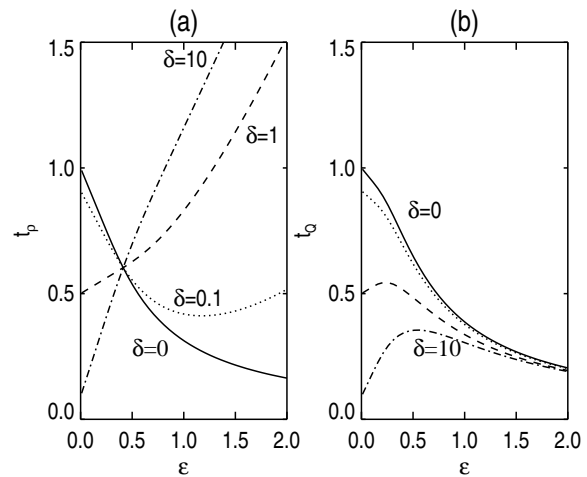}
\caption{ Analytical results for the exponential relaxation time from
  the linear theory for weak applied potentials, ($V \ll
  kT/ze$). The time scale for relaxation of the surface diffuse-charge
  density, $t_\rho$, from Eq.~(\protect\ref{eq:tcrho}) is shown in (a)
  and that of the total interfacial (half-cell) diffuse charge,
  $t_Q$, from Eq.~(\protect\ref{eq:tcQ}) in (b). In each case, the
  charging time, scaled to $\tau_c=L/\kappa D =\lambda_D L/D$, is
  plotted versus the dimensionless diffuse-layer thickness, $\epsilon
  = \lambda_D/L$, for different dimensionless Stern-layer thicknesses,
  $\delta = \lambda_S/\lambda_D = 0, 0.1, 1, 10$ (solid, dot, dash, and dot-dash lines, respectively).
  \label{fig:tc} }
\end{figure}

%%%%%%%%%%%%%%%%%%%%%%%%%%%%%%%%%
\section{ Dimensionless Formulation and Numerical Solution }
\label{sec:nondim}
%%%%%%%%%%%%%%%%%%%%%%%%%%%%%%%%%%%%%%%%%

\subsection{ Basic Equations }
\label{sec:eqns}

In preparation for analysis of the full,
nonlinear problem, we cast it in a dimensionless form using $L$ as the
reference length scale and $\tau_c=\lambda_D L/D$ as the reference
time scale, as motivated by the linear theory. Time and space are then
represented by $t=\tau D/\lambda_D L$ and $x=X/L$, and the problem
is reformulated through reduced variables: $c=(C_+ +C_-)/2C_0$ for the
local salt concentration, $\rho= (C_+-C_-)/2C_0=\rho_e/(2C_0ze)$ for
the charge density, and $\phi=ze\Phi/k_BT$ for the electrostatic
potential. The solution is determined by only three dimensionless
parameters: $v=zeV/k_BT$, the ratio of the applied voltage to the
thermal voltage, $\epsilon=\lambda_D/L$, the ratio of the Debye length
to the system size, and $\delta=\lambda_S/\lambda_D$, the ratio of the
Stern length to the Debye length~\cite{bonnefont01}.

With these definitions, the dimensionless equations for $-1 < x < 1$ and
$t>0$ are
\begin{eqnarray}
\frac{\partial c}{\partial t} &=& \epsilon\, \frac{\partial}{\partial x}\left(
\frac{\partial c}{\partial x} + \rho \frac{\partial \phi}{\partial x}\right)
\label{eq:c} \\ \frac{\partial \rho}{\partial t} &=& \epsilon\,
\frac{\partial}{\partial x}\left( \frac{\partial \rho}{\partial x} + c
\frac{\partial \phi}{\partial x}\right)  \label{eq:rho} \\ -\epsilon^2 \,
\frac{\partial^2 \phi}{\partial x^2} &=& \rho  \label{eq:phi}
\end{eqnarray}
with boundary conditions at $x = \pm 1$,
\begin{eqnarray}
\frac{\partial c}{\partial x} + \rho \frac{\partial \phi}{\partial x} &=& 0
\label{eq:bcc} \\ \frac{\partial \rho}{\partial x} + c \frac{\partial
\phi}{\partial x} &=& 0 \label{eq:bcrho} \\ v - \delta\,\epsilon\,
\frac{\partial \phi}{\partial x} &=& \pm\phi \label{eq:sternbc}
\end{eqnarray}
and initial conditions, $c(x,0)=1$, $\rho(x,0) = 0$, and $\phi(x,0) =
v\, x$. Note that the limit of a negligible screening length, 
$\epsilon\rightarrow 0$, is {\it singular} because it is impossible to
satisfy all the boundary conditions when $\epsilon=0$. Physically,
this corresponds to the limit of exact charge neutrality, $\rho=0$,
which is always violated to some degree at electrochemical interfaces.

The total diffuse charge near the cathode is
\begin{equation}
q(t) = \int_{-1}^0 \rho(x,t) dx ,
\end{equation}
scaled to $2zeC_0 L$. The dimensionless Faradaic current density is,
\begin{equation}
j_F = \frac{\partial \rho}{\partial x} + c \frac{\partial \phi}{\partial x}
,
\end{equation}
scaled to $2zeC_0D/L$ (Nernst's diffusion-limited
current~\cite{bonnefont01}).

\subsection{ Time Scales for Linear Response }

The time scale for exponential relaxation of the surface charge
density in the linear theory above, Eq.~(\ref{eq:taucrho}), has the
dimensionless form,
\begin{equation}
t_{\rho} = \frac{(1 + 3\delta\epsilon/2)\,\coth(\epsilon^{-1}) -
    \delta\,\csch^2(\epsilon^{-1})/2 - \epsilon }{
    1+\delta\,\coth(\epsilon^{-1}) } .  \label{eq:tcrho}
\end{equation}
As shown in Fig.~\ref{fig:tc}(a), this formula shows that for a wide
range of diffuse and Stern layer thicknesses, the basic time scale is
always roughly of order, $\lambda_D L/D$, since $t_c$ is of order $1$.  In the
limit of a thin diffuse double layer, the dimensionless time scale
has the  form,
\begin{equation}
t_{\rho} = \frac{1}{1+\delta} +   \left(\frac{3\delta -
  2}{2(1+\delta)}\right) \, \epsilon  + O(e^{-\epsilon^{-1}}).  \label{eq:tcrhothin}
\end{equation}
with exponentially small errors.  In the limit of a thin Stern layer,
the time scale becomes
\begin{eqnarray}
t_{\rho} &=& \coth(\epsilon^{-1}) - \epsilon + \nonumber \\
& &  \left[ 5\epsilon \coth(\epsilon^{-1}) - 2
\coth(\epsilon^{-1}) - \csch^2(\epsilon^{-1}) \right] \,
\frac{\delta}{2} \nonumber\\ & &  + \,
O(\delta^2) .
\end{eqnarray}
For simultaneously thin Stern and diffuse layers, we obtain the simple
result,
\begin{equation}
t_{\rho} \sim 1 - \epsilon - \delta
\end{equation}
which, as in Fig.~\ref{fig:tc}(a), shows that increasing either $\epsilon
= \lambda_D/L$ or $\delta = \lambda_S/\lambda_D$ tends to reduce the
charging time in this limit, compared to the leading-order value,
$\lambda_D L /D$. Putting the units back, this expression can be
written as,
\begin{equation}
\tau_{\rho} \sim \frac{\lambda_D L}{D} - \frac{\lambda_D^2}{D} -
\frac{\lambda_S L}{D}
\end{equation}
for $\lambda_S \ll \lambda_D \ll L$, which clearly shows the Debye
time, $\lambda_D^2/D$, appearing only as a small perturbation of the
intermediate time scale, $\lambda_D L/D$, for the relaxation of the cell.

\begin{figure*}
\includegraphics[width=5in]{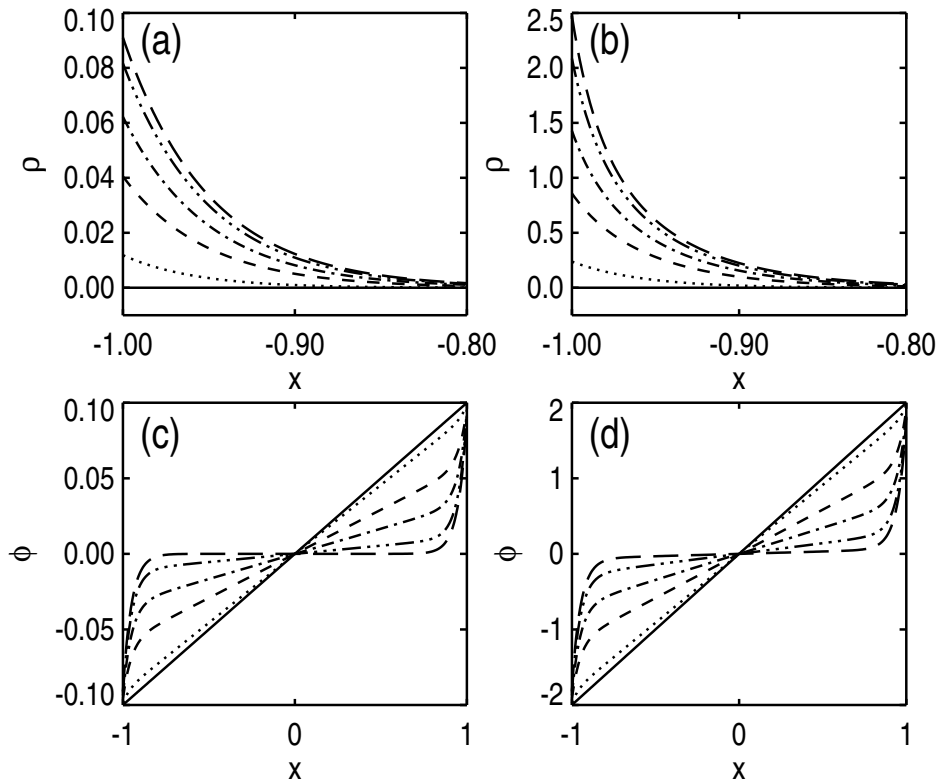}
\caption{ Profiles for $t=0$ (solid), $0.1$ (dot), $0.5$ (dash),
$1$ (dot-dash), $2$ (dot-dot-dot-dash), $\infty$ (long dash) of the
dimensionless charge density, $\rho(x,t)$, for dimensionless voltages
(a) $v=0.1$ and (b) $v=1$, and of the dimensionless potential, $\phi$,
for (c) $v=0.1$ and (d) $v=1$ ($\epsilon=0.05$,
$\delta=0.1$).
\label{fig:fields} }
\end{figure*}

Similar results hold for the relaxation time for the total half-cell
charge, Eq.~(\ref{eq:taucQ}), which has the dimensionless form,
\begin{equation}
t_Q = \frac{1 + \frac{1}{2}\sech^2(\epsilon^{-1}) +
  \frac{3\delta\epsilon}{2}}{\tanh(\epsilon^{-1}) + \delta} -
\frac{\epsilon}{2} - \frac{\sech(\epsilon^{-1})\tanh(\epsilon^{-1})}{2\left[1 -
    \sech(\epsilon^{-1})\right]}   \label{eq:tcQ}
\end{equation}
For thin double layers, we obtain the same leading-order behavior,
\begin{equation}
t_Q \sim \frac{1}{1+\delta} - \left[ \frac{1-2\delta}{2(1+\delta)}
  \right]\,\epsilon +  O(e^{-\epsilon^{-1}}) \label{eq:tcQthin}
\end{equation}
although the correction term is somewhat different for thick diffuse
layers. For simultaneously thin diffuse and Stern layers, the
dimensionless relaxation time for the total charge becomes,
\begin{equation}
t_Q \sim 1 - \frac{\epsilon}{2} - \delta .
\end{equation}
For a detailed summary of how the two time scales, $t_{\rho}$ and $t_Q$,
depend on the parameters, $\epsilon$ and $\delta$, see Figure \ref{fig:tc} (a)
and (b), respectively.

\subsection{ Numerical Solution}
%\label{sec:num}

Our dimensionless model problem, stated in Section ~\ref{sec:eqns}, is
straightforward to solve numerically using finite differences, at
least if $\epsilon$ is not too small. (Ironically, as shown below,
analytical progress is much easier in this singular limit.)  To
resolve the boundary layer where the gradient is large, a variable
size mesh is used, along with second-order-accurate differencing that
accounts for the variable grid sizes.  The third-order Adams-Bashforth
method is used in time.  The number of the grid points and the 
ratio of the smallest to largest grid size are
varied depending on the values of $\epsilon$ and $v$.  
The numerical convergence is verified though multiple runs of
different resolutions, and as a result, up to 1024 points are used
in calculations for higher $v$.

To maximize the importance of diffuse charge,
we first consider a rather larger value of $\epsilon$, even for a
micro-electrochemical system, $\epsilon = 0.05$, say for $\lambda_D =
5$ nm and $L = 0.1 \mu$m. The Stern length is always of molecular
dimensions, so we choose $\lambda_S = 5$ \AA, and thus $\delta =
0.1$. The time evolutions of the charge and potential are shown in
Fig.~\ref{fig:fields} for $v = 0.1$ and $v=1$. At room temperature
(and $z=1$), these voltages correspond to $V = 2.5$ mV and
$V=25$ mV, respectively, which, when transferred to the diffuse layer
after screening give maximum electric fields of order $10$ V/$\mu$m.

\begin{figure}
\includegraphics[width=\linewidth]{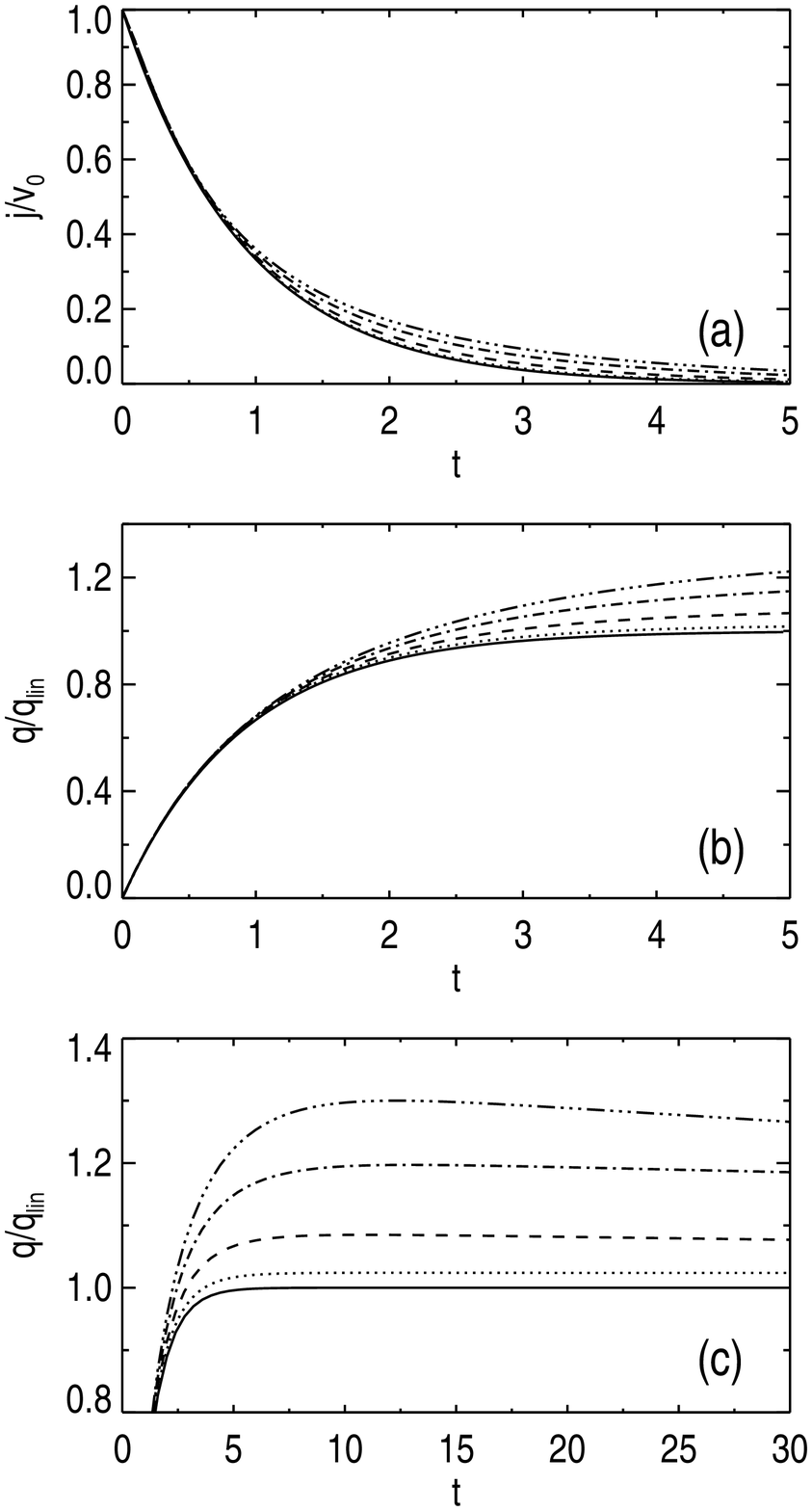}
\caption{ (a) The dimensionless current density, $j(t)$ (in units of
$2zeC_0 D/L$), and (b) the dimensionless total diffuse charge on the
cathodic side of the cell, $q(t)$ (in units of $2 z e C_0 L$), scaled
to $q_o = v/(1+\delta)$, versus dimensionless time, $t$ (in units of
$\tau_c = \lambda_D L/ D$). Numerical results for dimensionless
voltages, $v= 1$ (dot), $2$ (dash), $3$ (dot-dash), and $4$
(dot-dot-dot-dash) are compared with linear dynamics in the thin
double-layer limit: $q(t)/q_o \sim 1 - e^{-(1+\delta) t}$ and $j(t)/v
\sim e^{-(1+\delta) t}$ (solid lines) as $v, \epsilon \rightarrow 0$.
The breakdown of linear theory for $v \geq 1$ is highlighted in (c),
where the data in (b) is replotted for longer times. \label{fig:q} }
\end{figure}

The current, $j$, and the total cathodic diffuse charge, $q$, are
plotted versus time, $t$, in Fig.~\ref{fig:q} for applied voltages, $v
= 1,2,3,$ and $4$. In all cases, the linearization is accurate at early
times ($t<1$) since the dimensionless voltage across the diffuse layer
remains small ($<1$). For $v=1$, the linear approximation is
reasonable for all times, but for somewhat larger voltages, $v= 2, 3,$ and $4$,
the relative error becomes unacceptable at long times, $t > 1$. 
%I removed a paragraph change here --K 
Not only is the limiting value of the total charge significantly
underestimated, but the dynamics also continues for a longer time,
with a qualitatively different charging profile.  The largest applied
voltage, $v=4$, shows this effect most clearly, as there is a
secondary relaxation at a much larger time scale of order $t =
1/\epsilon = 20$. Unlike the other cases, which display the expected
steady increase in charge of an RC circuit, for $v\geq 4$ the total
charge quickly reaches a maximum value, after the initial RC charging
process, and then slowly decays toward its limiting value.

We are not aware of any previous theoretical prediction of such a
non-monotonic charging profile, so it is a major focus of this work
(in sections ~\ref{sec:higher} and \ref{sec:strong}). It is
reminiscent of the Warburg impedance due to bulk diffusion of
current-carrying ions at the time scale, $\tau_L = L^2/D$, or
$1/\epsilon$ in our units, in (linear) response to Faradaic processes,
which consume or produce them at an electrode. Here, however, there
are no Faradaic processes, so any such bulk diffusion must be related
to the adsorption or desorption of ions in the diffuse part of the
double layer. Moreover, the over-relaxation of the charge density is
part of the {\it non-linear} response to a large applied voltage, so
it will require more sophisticated analytical methods.

%%%%%%%%%%%%%%%%%%%%%%%%%%%%%%%%%
\section{ Weakly Nonlinear Dynamics }
\label{sec:nonlin}
%%%%%%%%%%%%%%%%%%%%%%%%%%%%%%%%%%%%%%%%%

\subsection{ Asymptotic Analysis for Thin Double Layers }

The remarkable robustness of the charging time well into the nonlinear
regime (at least for the primary relaxation phase) can be predicted analytically using matched
asymptotic expansions in the singular limit of thin double layers,
$\epsilon = \lambda_D/L \ll 1$. Most (if not all) previous studies of
time-dependent problems using asymptotic analysis have scaled time to
the diffusion time, $\tau_L = L^2/D$. In this section, we will see how
the correct charging time scale, $\tau_c = \lambda_D L/D$, arises
systematically from asymptotic matching at leading order.  We also
consider, apparently for the first time, the general case of arbitrary
voltage, $v= ze V/k_BT$, and Stern-layer thickness, $\delta =
\lambda_S/\lambda_D$, with a time-dependent zeta-potential (i.e. potential drop
over the diffuse layer). We also
study higher-order corrections, which involve some bulk diffusion at
the time scale, $\tau_L$.

As usual, matched asymptotic expansions only produce a series of
``asymptotic'' approximations to the solution, in the sense that
higher terms in the expansions vanish more quickly than the leading
terms as $\epsilon \rightarrow 0$, with the other parameters, $v$ and
$\delta$, held fixed at arbitrary values. For any fixed $\epsilon >0$
(no matter how small), there could be $\epsilon$-dependent
restrictions on $v$ and $\delta$ for various truncated expansions to
produce accurate approximations. We refer to the regime where such
conditions hold as ``weakly nonlinear'', as opposed to the ``strongly
nonlinear'' regime where the asymptotic expansions break down
(described below in section~\ref{sec:strong}).

\subsection{ Outer and Inner Expansions }

We begin by seeking regular asymptotic expansions (denoted by a bar
accent) in the bulk ``outer'' region, e.g.
\begin{equation}
c(x,t) \sim \bar{c}(x,t) = \bar{c}_0 + \epsilon \, \bar{c}_1 +
\epsilon^2\, \bar{c}_2 + \ldots .   \label{eq:cout}
\end{equation}
Substituting such expansions into Eqs.~(\ref{eq:c})--(\ref{eq:phi})
and equating terms order by order yields a hierarchy of
partial-differential equations. At leading order in $\epsilon$, we
find that the bulk concentration does not vary in time, $\bar{c}_0 = 
1$, simply because the charging time scale, $\tau_c$, is much {\it
smaller} than the bulk diffusion time scale, $\tau_L$. The
leading-order potential is linear, 
\begin{equation}
\bar{\phi}_0 = \bar{j}_0(t)\,  x,  \label{eq:phi0}
\end{equation}
where $\bar{j}_0(0) = v$. Since the leading-order bulk concentration is
uniform, $\bar{j}_0(t)$ is the leading-order current density. The
leading-order charge density,
\begin{equation}
\bar{\rho}_2 = -\frac{\partial^2 \bar{\phi}_0}{\partial x^2}
\end{equation}
vanishes because the leading-order potential, Eq.\ (\ref{eq:phi0}), is harmonic, although
at next order, $O(\epsilon^3)$, a nonzero
bulk charge density, $\bar{\rho}_3$, arises due to concentration polarization. (See below.)
These arguments justify the usual assumption of bulk electroneutrality
to high accuracy, even during interfacial charging, as long as the
dynamics are ``weakly nonlinear''.

The regular outer approximations must be matched with singular ``inner''
approximations in the boundary layers.  The problem has the following
symmetries about the origin,
\begin{equation}
\begin{array}{c}
c(-x,t) = c(x,t) \\ 
\rho(-x,t) = -\rho(x,t) \\
 \phi(-x,t) = -\phi(x,t)
\end{array}
 \label{eq:sym}
\end{equation}
so we consider only the boundary layer at the cathode,
$x=-1$, by transforming the equations to the inner coordinate, $y =
(x+1)/\epsilon$:
\begin{eqnarray}
\epsilon \, \frac{\partial \tilde{c}}{\partial t} &=&
\frac{\partial}{\partial y}\left( \frac{\partial \tilde{c}}{\partial y} +
\tilde{\rho} \frac{\partial \tilde{\phi}}{\partial y}\right) \label{eq:cin}
\\ \epsilon \, \frac{\partial \tilde{\rho}}{\partial t} &=&
\frac{\partial}{\partial y}\left( \frac{\partial \tilde{\rho}}{\partial y}
+ \tilde{c} \frac{\partial \tilde{\phi}}{\partial y}\right)
\label{eq:rhoin}
\\ -\frac{\partial^2 \tilde{\phi}}{\partial y^2} &= &\tilde{\rho} \label{eq:phiin}
\end{eqnarray}
This scaling removes the singular perturbation in
Poisson's equation, so we can seek regular asymptotic expansions for the
inner approximations (denoted by tilde accents), e.g.
\begin{equation}
c(x,t) \sim \tilde{c}(y,t) = \tilde{c}_0 + \epsilon \, \tilde{c}_1 +
\epsilon^2 \, \tilde{c}_2 + \ldots
\end{equation}
Matching with the bulk approximations {\it in
space} involves the usual van Dyke conditions, e.g.
\begin{equation}
\lim_{y\rightarrow\infty} \tilde{c}(y,t) \sim \lim_{x\rightarrow -1}
\bar{c}(x,t) ,
\end{equation}
which implies $\tilde{c}_0(\infty,t) = \bar{c}_0(-1,t)$,
$\tilde{c}_1(\infty,t) = \bar{c}_1(-1,t)$, etc., but we will also have
to make sure that the expansions are properly synchronized {\it
in time}. In particular, we will have to worry about the appearance of
multiple time scales at different orders.

Substituting the inner expansions into the rescaled 
equations ~(\ref{eq:cin})--(\ref{eq:phiin}) causes the time-dependent terms
to drop out at leading order.  Physically, this quasi-equilibrium
occurs because the charging time, $\tau_c$, is much {\it larger} than
the Debye time, $\tau_D$, characteristic of local dynamics in the
boundary layer (at the scale of the Debye length, $\lambda_D$). As a
result, we systematically arrive at classical Gouy-Chapman profiles
for the equilibrium diffuse layer at leading order,
\begin{equation}
\tilde{c}_\pm \sim e^{\mp \tilde{\psi}}, \ \ \tilde{c}_0 = \cosh
\tilde{\psi}_0 , \ \ \tilde{\rho}_0 = - \sinh \tilde{\psi}_0
\end{equation}
where the excess voltage relative to the bulk, 
\begin{equation}
\tilde{\psi}(y,t) =
\tilde{\phi}(y,t) - \bar{\phi}(-1,t) \sim \tilde{\psi}_0 + \epsilon\, 
\tilde{\psi}_1 + \ldots,
\end{equation}
satisfies the Poisson-Boltzmann equation at leading order,
\begin{equation}
\frac{\partial^2 \tilde{\psi}_0}{\partial y^2} = \sinh \tilde{\psi}_0 .
\end{equation}
Note that matching implies $\tilde{\psi}_0(\infty,t) =
\tilde{\psi}_1(\infty,t)=\ldots = 0$. The dimensionless zeta
potential, $\tilde{\zeta}(t) = \tilde{\psi}(0,t)$, varies as the
diffuse layer charges.

After the first integration we apply matching to the electric
field,
\begin{equation}
\frac{\partial\tilde{\phi}}{\partial y}(\infty,t) \sim \epsilon \, 
\frac{\partial\bar{\phi}}{\partial x}(-1,t) \ \longrightarrow \ 
\frac{\partial\tilde{\phi}_0}{\partial y}(\infty,t) = 0,
\label{eq:Ematch}
\end{equation}
to obtain 
\begin{equation}
\frac{\partial \tilde{\psi}_0}{\partial y} = - 2 \, \sinh(\tilde{\psi}_0/2).
\label{eq:psiy}
\end{equation}
After the second integration,
\begin{equation}
\tilde{\psi}_0(y,t) = - 4\, \tanh^{-1}(e^{-(y+K(t))}),   \label{eq:psiin0}
\end{equation}
we are left with a constant, 
\begin{equation}
K(t) = \log\coth (-\tilde{\zeta}_0(t)/4),   \label{eq:K}
\end{equation}
to be determined from $\tilde{\zeta}_0(t)$ (below) by the Stern
boundary condition at the cathode surface, $y=0$, and the coupling to
the bulk region.  The offset parameter, $K(t)$, which also appears in
the concentration and charge density,
\begin{eqnarray}
\tilde{c}_0(y,t) &= & 1 + 2\, \csch^2(y + K(t)) \label{eq:cin0} \\ 
\tilde{\rho}_0(y,t) &= & 2\, \csch(y+K(t))\coth(y+K(t)) \label{eq:rhoin0}
\end{eqnarray}
is quite sensitive to Faradaic reactions~\cite{bonnefont01}, but here
we focus only on the effect of compact-layer capacitance.

\subsection{ Time-dependent Matching }

It seems we have reached a paradox: Both the bulk and the boundary
layers are in quasi-equilibrium at leading order, and yet there must
be some dynamics, if we have chosen the proper time scale. The
resolution lies in taking a closer look at asymptotic
matching. Physically, we are motivated to consider the dynamics of the
total diffuse charge, which has the scaling, $q(t) \sim \epsilon\, \tilde{q}(t)
$, where
\begin{equation}
\tilde{q} = \int_0^\infty \tilde{\rho}(y,t)\, dy \sim \tilde{q}_0
+ \epsilon\, \tilde{q}_1 + \epsilon^2 \, \tilde{q}_2  + \ldots \label{eq:qt}
\end{equation}
Taking a time derivative using Eq.~(\ref{eq:rhoin}) and applying the no-flux
boundary condition (\ref{eq:bcrho}), we find
\begin{equation}
\frac{d \tilde{q}}{dt} = \lim_{y \to \infty} \frac{1}{\epsilon}
\left(\frac{\partial \tilde{\rho}}{\partial y} + \tilde{c} \frac{\partial
\tilde{\phi}}{\partial y}\right)   
 \sim \lim_{x\to -1} \left(\frac{\partial
\bar{\rho}}{\partial x} + \bar{c} \frac{\partial \bar{\phi}}{\partial
x}\right)
\label{eq:qmatch}
\end{equation}
where we have applied matching to the {\it derivatives} (flux
densities).  Substituting the inner and outer expansions yields a
hierarchy of matching conditions. At leading order, we have
\begin{equation}
\frac{d\tilde{q}_0}{dt}(t) = \bar{j}_0(t),
  \label{eq:dq}
\end{equation}
which shows that we have chosen the right time scale because this is a
balance of O(1) quantities.  Moreover, it can be shown that any other
choice of scaling would lead to a breakdown of asymptotic matching in
the limit $\epsilon \rightarrow 0$. (For example, in the analogous
Equations (42)--(43) of Ref.~\cite{gonzalez00} for small AC
potentials, the time-dependent term vanishes in this limit, showing
that the proper scaling was not used.) Therefore, the correct charging
time scale, Eq.~(\ref{eq:tmixed}), in the weakly nonlinear regime
follows systematically from time-dependent asymptotic matching at
leading order.

The physical interpretation of Equation~(\ref{eq:dq}) is clear: At
leading order, the boundary layer acts like a capacitor, whose
total charge (per unit area), $\tilde{q}$, changes in response to the
transient Faradaic current density, $\bar{j}(t)$, from the bulk.  The
matching condition can also be understood physically as a statement of
current continuity across the diffuse
layer. Substituting Poisson's equation~(\ref{eq:phiin}) into
Eq.~(\ref{eq:qt}), integrating, and matching the electric field using
Eq.~(\ref{eq:Ematch}), we see that the left hand side of
Eq.~(\ref{eq:dq}) is simply the leading-order (dimensionless) {\it
displacement current
density}~\cite{bonnefont01,bonnefont_thesis,brumleve78,cohen65} at the
cathode surface,
\begin{equation}
\frac{d \tilde{q}_0}{dt} = \frac{\partial}{\partial t} \frac{\partial
\tilde{\phi}_0}{\partial y}(0,t) = \tilde{j}_0(t) ,
\end{equation}
so the matching condition simply reads, $\tilde{j}_0(t) =
\bar{j}_0(t)$. This transient displacement current exists in the
external circuit, even if there is no Faradaic current.

\subsection{ Leading-order Dynamics }

Using Eqs.~(\ref{eq:phiin}), (\ref{eq:Ematch}) and (\ref{eq:psiy}),
the integral in Eq.~(\ref{eq:qt}) can be performed at leading order to
obtain the Chapman's formula for the total diffuse charge,
\begin{equation}
\tilde{q}_0 = - 2\, \sinh(\tilde{\zeta}_0/2) . \label{eq:qzeta}
\end{equation}
The Stern boundary condition, Eq.~(\ref{eq:sternbc}), then yields,
\begin{equation}
%\Delta\phi_D + 2\, \delta\, \sinh(\Delta\phi_D/2) = \Delta\phi_i ,
\tilde{\zeta}_0 + 2\, \delta\, \sinh(\tilde{\zeta}_0/2) = \bar{j}_0(t)
- v = \tilde{\Psi}_0 ,
\label{eq:phid}
\end{equation}
where $\tilde{\Psi}(t) = - v - \bar{\phi}(-1,t) \sim \tilde{\Psi}_0 +
\epsilon \tilde{\Psi}_1 + \ldots$ is the total voltage across the
compact and diffuse layers.  Substituting into the matching condition,
Eq.~(\ref{eq:dq}), we obtain an ordinary, initial-value problem,
either for the leading-order double-layer voltage
% ($ \Delta\phi_i(t)\rightarrow v$),
\begin{equation}
-\tilde{C}_0(\tilde{\Psi}_0)\, \frac{d\tilde{\Psi}_0}{dt} =
\tilde{\Psi}_0 + v,\ \ \ \tilde{\Psi}_0(0)=0, \label{eq:PsiODE}
\end{equation}
or for the leading-order current density
% ($j(t) \rightarrow 0$),
\begin{equation}
\tilde{C}_0(\bar{j}_0-v)\, \frac{d\bar{j}_0}{dt} = -\bar{j}_0 , \ \ \
\bar{j}_0(0) = v,  \label{eq:jODE}
\end{equation}
where $\tilde{C}_0(\tilde{\Psi}_0) = d\tilde{q}_0/d\tilde{\Psi}_0$ is
the differential capacitance for the double layer as a function of its
total voltage, relative to the potential of zero charge. 

\begin{figure}
\includegraphics[width=2.5in]{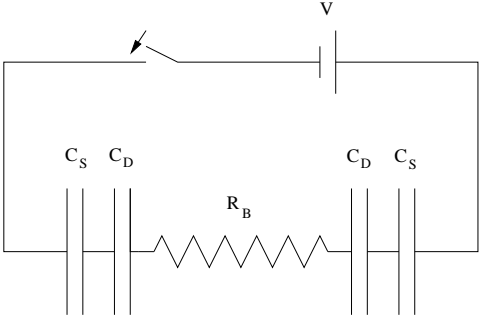}
\caption{ Sketch of the equivalent RC circuit for the leading-order
weakly nonlinear approximation: compact-layer and diffuse-layer
capacitors in series with a bulk resistor. Although remarkably robust,
the circuit approximation is violated by higher-order corrections,
especially at large voltages.
\label{fig:RC}}
\end{figure}

The effective double-layer capacitance is given by
\begin{equation}
\tilde{C}_0 = \frac{1}{\mbox{sech}(\tilde{\zeta}_0/2) + \delta}   \label{eq:Ci}
\end{equation}
where $\tilde{\zeta}_0$ is related to $\tilde{\Psi}_0$ by
Eq.~(\ref{eq:phid}). A similar formula arises in the classical
circuit model of Macdonald~\cite{macdonald54}. Indeed, the
leading-order charging dynamics from asymptotic analysis corresponds
exactly to the nonlinear RC circuit shown in Fig. ~\ref{fig:RC}. We
expect, however, that the {\it ad hoc} circuit approximation cannot
describe higher-order asymptotic approximations, where the finite
thickness of the double layer becomes important.

Linearizing for small voltages, $\tilde{C}_0 \sim
1/(1+\delta)$, we obtain the same results as before in the limit
$\epsilon\rightarrow 0$, now by a completely different method,
\begin{eqnarray}
\bar{j}_0(t) &\sim& v\, e^{-(1+\delta) t} = v + \tilde{\Psi}_0(t) \\
\tilde{q}_0(t) &\sim& \frac{v (1 - e^{-(1+\delta) t})}{1+\delta}  
\end{eqnarray}
As shown in Fig.~\ref{fig:q} for $\delta=0.1$, the linearization
describes the charging dynamics fairly accurately, even for somewhat large
voltages ($v \approx 1$), as long as $\delta$ is not too small.  One way to understand
this is that the total differential capacitance satisfies the uniform
bounds,
\begin{equation}
\frac{1}{1+\delta} = \tilde{C}_0(0) \leq \tilde{C}_0(\tilde{\Psi}_0) <
\tilde{C}_0(\infty) = \frac{1}{\delta} , \label{eq:Cbound}
\end{equation}
in the linear and nonlinear regimes. Moreover, the linearization is
always accurate at early times (up to $t \approx 1$ or $\tau \approx
\tau_c$) for any applied voltage, as long as the initial
zeta potential (or diffuse charge) is small. This is also clearly seen
in Fig.~\ref{fig:q}.

\begin{figure}
\includegraphics[width=3in]{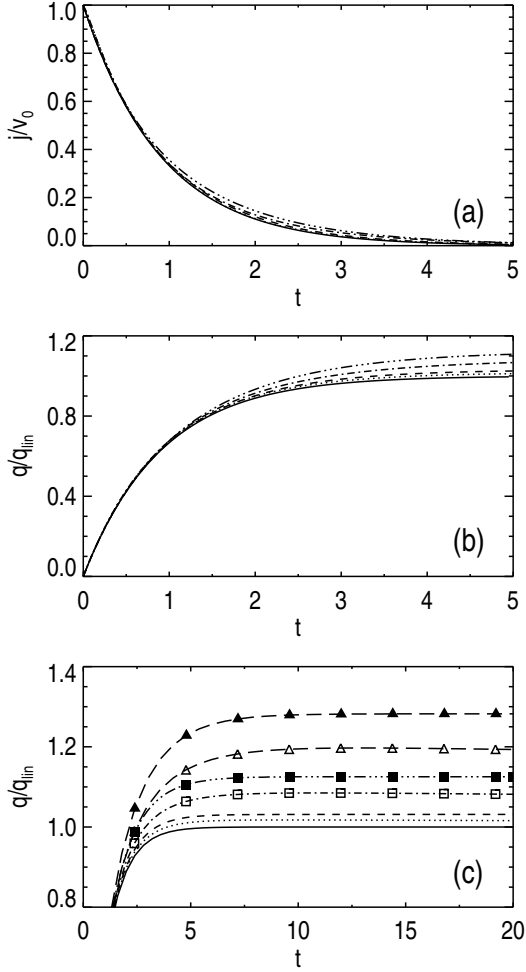}
\caption{ The comparison of the full numerical solution with the
leading-order asymptotic results: (a) $j/v$, (b) $q/q_{lin}$ (early
evolution), and (c) $q/q_{lin}$ (long time evolution).  The full
numerical solution are shown with dot ($v=1$), dot-dash (with open squares in (c)) ($v=2$) and
long dash with open triangles ($v=3$).  The leading-order asymptotic results are
plotted with dash ($v=1$), dot-dot-dot-dash (with filled squares in (c)) ($v=2$) and long dash with filled triangles
($v=3$).  The curves for $v=3$ are omitted in (a) and (b) for clarity.
The solid lines show the linear dynamics in the thin double-layer
limit.  \label{fig:qa}}
\end{figure}

The dynamical equation (\ref{eq:PsiODE}) or (\ref{eq:jODE}) is
first-order and separable, so its exact solution is easily expressed
in integral form,
\begin{equation}
\tilde{\Psi}_0(t) = \bar{j}_0(t) - v = - F^{-1}(t)  \label{eq:jexact}
\end{equation}
where 
\begin{equation}
F(z)  = \int_0^z \frac{\tilde{C}_0(u)\,du}{u+v}
\label{eq:F}
\end{equation}
The integral can be evaluated numerically and the total charge
recovered from the Eqs.~(\ref{eq:qzeta}) and (\ref{eq:phid}). The
results in Fig.~\ref{fig:qa} show that the leading-order dynamics
compares fairly well with the numerical solution to the full nonlinear
problem for $\epsilon=0.05$ and $\delta=0.1$, at least for the decay
of the current density, especially at early times ($t \approx 1$). The
limiting value of the total diffuse charge is also approximated much
better than in the linear theory (Fig.~\ref{fig:q}), due to the
nonlinear differential capacitance, Eq.~(\ref{eq:Ci}). For large
voltages ($v > 1$), however, total charge shows some secondary
dynamics at longer time scales ($t \gg 1$), which is not fully
captured by the leading-order asymptotic approximation (or the
corresponding circuit model). As we shall see below, this can only be
understood by considering higher-order terms which violate the circuit
approximation.

% CONVERGENCE TEST
%\begin{figure}
%\includegraphics[width=3in]{figconv.eps}
%\caption{
%\label{fig:figconv} }
%\end{figure}

For moderately large voltages ($v \approx 1$), we can expand around
$u=v$ in the integrand of Eq.~(\ref{eq:F}) and obtain a
long-time exponential decay,
\begin{equation}
\bar{j}_0(t)  = v + \tilde{\Psi}_0(t) \propto e^{- t/\tilde{C}_0(v)} 
\label{eq:jlate}
\end{equation}
as $t \rightarrow \infty$. This reveals a (dimensionless)
characteristic time, $t_c = \tilde{C}_0(v)$, which is larger than that
of the linear regime, $t_c = \tilde{C}_0(0) = 1/(1+\delta)$, by at most
a factor of $1+1/\delta$ ($=11$ in our numerical examples). Although
this factor is non-negligible, the characteristic time, $\tau_c$, is
still the basic time scale, rather than $\tau_L$ and $\tau_D$ which differ
from $\tau_c$ by factors of $\epsilon$, i.e. usually two or more
orders of magnitude. As the voltage is increased, however,
nonlinearity always becomes important, and one of its generic effects
is to slow down the relaxation process.

In order to simplify the response function, $F(z)$, and other
quantities, it is useful to consider the regular limit of thin Stern
layers, $\delta\rightarrow 0$ (taken after the singular limit of thin
diffuse layers, $\epsilon \rightarrow 0$). In this common physical
regime, where $\lambda_S \ll \lambda_D \ll L$, the following
asymptotic expansions can be derived by iteration~\cite{hinch} from
Eqs.~(\ref{eq:qzeta}), (\ref{eq:phid}), and (\ref{eq:Ci}):
\begin{eqnarray}
\tilde{\zeta}_0 &\sim& \tilde{\Psi}_0 - 2\delta \sinh\frac{\tilde{\Psi}_0}{2}
+ \delta^2 \sinh \tilde{\Psi}_0 + \ldots \label{eq:phida} \\ 
\tilde{q}_0 &\sim& - 2 \sinh\frac{\tilde{\Psi}_0}{2} 
+ \delta \sinh \tilde{\Psi}_0 \nonumber\\ &
& - \delta^2 \left( \sinh \tilde{\Psi}_0 \cosh\frac{\tilde{\Psi}_0}{2} +
\sinh^3\frac{\tilde{\Psi}_0}{2} \right)
\label{eq:qa} \\ 
\tilde{C}_0 &\sim& \cosh\frac{\tilde{\Psi}_0}{2} 
- \delta \cosh \tilde{\Psi}_0  \nonumber \\ & & +
\delta^2 \left( \cosh \tilde{\Psi}_0\, \cosh\frac{\tilde{\Psi}_0}{2} +
\frac{1}{2}\, \sinh \tilde{\Psi}_0 \, \sinh\frac{\tilde{\Psi}_0}{2}
\right. \nonumber \\
& & \left. + \frac{3}{2}\, \sinh^2\frac{\tilde{\Psi}_0}{2}\,
\cosh\frac{\tilde{\Psi}_0}{2} \right)  \label{eq:Ca} 
\end{eqnarray}
The response function can then be expanded in somewhat simpler (but
still nontrivial) integrals,
\begin{equation}
F(z) \sim \int_0^z \frac{\cosh(u/2)\, du}{u+v} - \delta \int_0^z
\frac{\cosh(u)\, dx}{u+v} + \ldots
\end{equation}
in the limit $\delta \rightarrow 0$.

%\begin{figure}
%\includegraphics[width=3in]{fig5.eps}
%\caption{ For $\delta=0.1$, plots of (a) the diffuse-layer voltage,
%$\Delta\phi_D$, and (b) the differential total capacitance, each
%versus the dimensionless interfacial voltage, $-\tilde{\Psi}_0$ (dotted
%lines).  (All quantities are dimensionless.) The corresponding points
%test the asymptotic approximations, Eqs.~(\protect\ref{eq:phida}) and
%(\protect\ref{eq:Ca}), to first (cross) and second order (diamond) in
%$\delta$, respectively.\label{fig:C} }
%\end{figure}

\begin{figure*}
\includegraphics[width=6in]{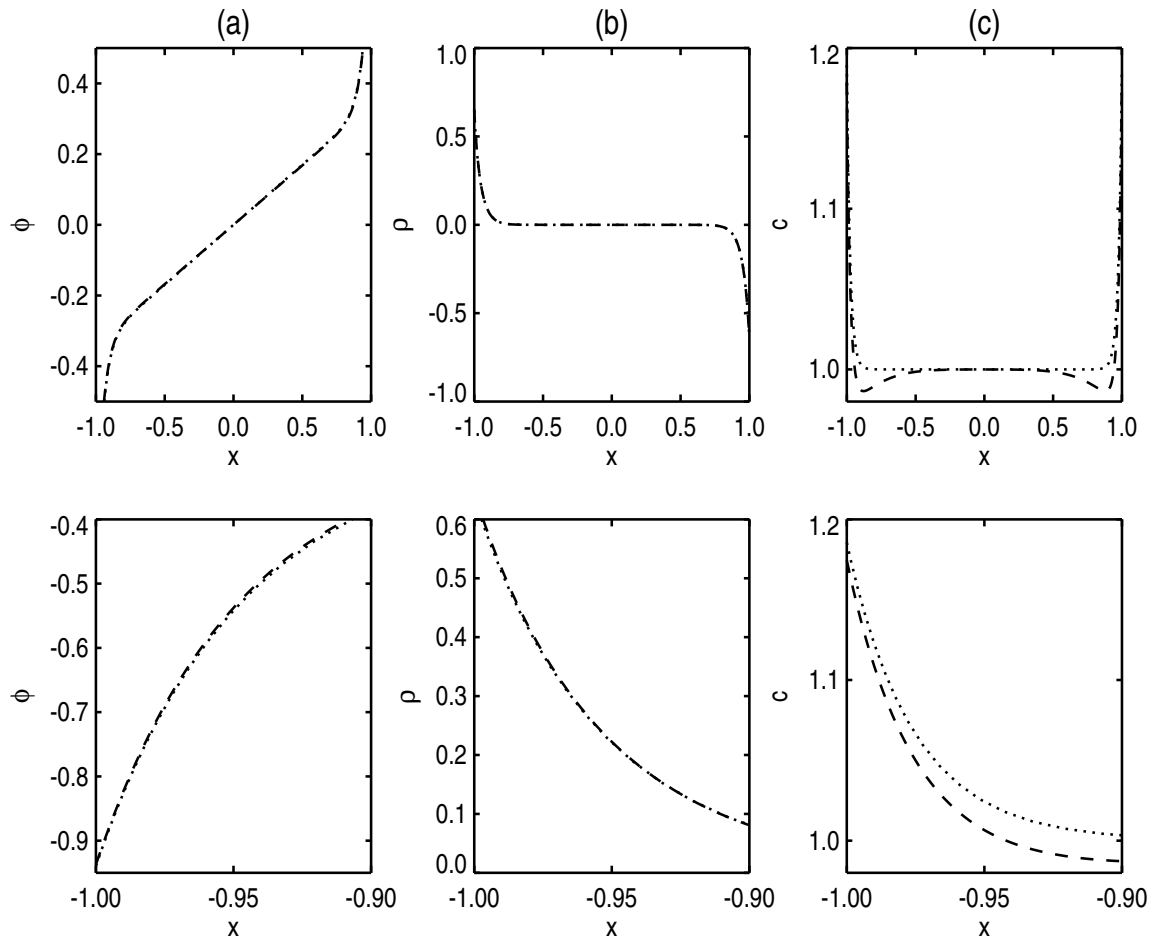}
\caption{ The potential (a), charge density (b) and concentration (c)
at $t=1$ for a large dimensionless voltage, $v=1$, with
$\epsilon=0.05$ and $\delta=0.1$. The full numerical solution (dashed 
lines) is compared with the leading-order uniformly valid
approximation (dotted lines),
Eqs.~(\ref{eq:uphi})--(\ref{eq:urho}).  The analytical 
approximations are almost indistinguishable from the numerical
solutions for $\phi$ and $\rho$, but not for $c$, which shows
errors of a few percent ($< \epsilon$) just outside the double layers.
\label{fig:compare} }
\end{figure*}

\subsection{ Uniformly Valid Approximations }

Asymptotic analysis tells us not only the behavior of integrated
quantities like total charge and voltage, but also the complete
spatio-temporal profiles of the charge density and potential. As
usual, uniformly valid approximations (in space) are constructed by
adding the outer and inner approximations and subtracting the
overlaps. Taking advantage of the symmetries in Eq.~(\ref{eq:sym}), we
obtain the following leading-order approximations:
\begin{eqnarray}
\phi(x,t) &\sim& \overline{j}(t)\, x +
\tilde{\psi}_0\left(\frac{1+x}{\epsilon},t\right) - 
\tilde{\psi}_0\left(\frac{1-x}{\epsilon},t\right) \label{eq:uphi}  \\ c(x,t)
&\sim& \tilde{c}_0\left(\frac{1+x}{\epsilon},t\right) +
\tilde{c}_0\left(\frac{1-x}{\epsilon},t\right) - 1 \label{eq:uc} \\
\rho(x,t) &\sim& \tilde{\rho}_0\left(\frac{1+x}{\epsilon},t\right) -
\tilde{\rho}_0\left(\frac{1-x}{\epsilon},t\right) \label{eq:urho}
\end{eqnarray}
where the boundary-layer contributions are given by
Eqs.~(\ref{eq:psiin0})--(\ref{eq:rhoin0}) and Eq.~(\ref{eq:phid}),
which express the effect of the compact layer. The time dependence of
the leading-order approximations is entirely determined by the bulk
current density, $\bar{j}_0(t)$, or the double-layer voltage,
$\tilde{\Psi}_0(t)$, via Eqs.~(\ref{eq:jexact}) and (\ref{eq:F}).

As shown in Fig.~\ref{fig:compare}, the time-dependent approximations
for $\phi$ and $\rho$ are in excellent agreement with our numerical
results well into the nonlinear regime ($v=1$), even for a fairly
large boundary-layer thickness, $\epsilon = 0.05$. The charge density
clearly shows the expected separation into three regions: a neutral
bulk with two charged boundary layers of $O(\epsilon)$ width.  On the
other hand, for the same parameters, the leading-order approximation
of $c$ is not nearly as good. As expected, the concentration exhibits
a homogeneous bulk region and two inhomogeneous boundary layers of
$O(\epsilon)$ width, which are fairly well described, but there are
also intermediate regions of depleted concentration extending far into
the bulk, which are not captured at leading order.

\section{ Higher-Order Effects }
\label{sec:higher}

\subsection{ Neutral-Salt Adsorption by the Double Layer }
 
We have seen that each diffuse-charge layer acquires an excess salt 
concentration relative to the outer region. At leading order, however,
there is no sign of how the extra ions got there.  This paradox, which
also applies to circuit models, is apparent from symmetry alone,
Eq.~(\ref{eq:sym}) --- Diffuse charge near the cathode grows by
bulk electromigration, which creates equal and opposite diffuse charge
near the anode. In contrast, the excess concentration is the same in
both double layers, so it can only arrive there by {\it diffusion} of
neutral electrolyte from the bulk, which is excluded at leading order.

The key to understanding higher-order terms, therefore, is the total
excess concentration per unit surface area in (say) the cathodic
diffuse layer, $w(t) = \epsilon\, \tilde{w}(t)$, where
\begin{equation}
\tilde{w}(t) = \int_0^\infty \left[\tilde{c}(y,t)-\bar{c}_0(-1,t)\right]
dy = \tilde{w}_0(t) + \epsilon\, \tilde{w}_1(t) + \ldots   \label{eq:w}
\end{equation}
is analogous to the scaled total diffuse charge, $\tilde{q}(t)$. (Note
that $\bar{c}_0(-1,t)=1$ in our model problem, but Equation
~(\ref{eq:w}) is more general.)  We proceed with matching in the same
manner as above.  Taking a time derivative using Eq.~(\ref{eq:cin})
and applying the no-flux boundary condition (\ref{eq:bcc}), we find
\begin{equation}
\frac{d \tilde{w}}{dt} = \lim_{y \to \infty} \frac{1}{\epsilon}
\left(\frac{\partial \tilde{c}}{\partial y} + \tilde{\rho} \frac{\partial
\tilde{\phi}}{\partial y}\right)   
 \sim \lim_{x\to -1} \left(\frac{\partial
\bar{c}}{\partial x} + \bar{\rho} \frac{\partial \bar{\phi}}{\partial
x}\right)
\label{eq:wmatch-full}
\end{equation}
Substituting the inner and outer expansions yields another 
hierarchy of matching conditions. At leading order, we have
\begin{equation}
\frac{d\tilde{w}_0}{d\bar{t}}(t)
= \frac{1}{\epsilon}\, \frac{d\tilde{w}_0}{dt}(t) = \frac{\partial
  \bar{c}_1}{\partial x}(-1,t) ,
  \label{eq:wmatch}
\end{equation}
which, unlike Eq.~(\ref{eq:dq}), involves a {\it new time variable},
\begin{equation}
\bar{t} ={\epsilon} t = \frac{\epsilon\,\tau}{\tau_c} =
\frac{\tau}{\tau_L} ,
%%%% QQQ Martin, please make sure my changes are OK above
% What did you do?
\end{equation}
scaled
to the bulk diffusion time, $ \tau_L=L^2/D$. Physically, this matching
condition simply expresses mass conservation: The (zeroth order)
excess concentration in the diffuse layer varies in response to the
(first order) diffusive flux from the bulk.

In Eq.~(\ref{eq:w}), the left-hand side is given by the leading-order
inner approximation calculated above.  Substituting
Eq.~(\ref{eq:cin0}) into Eq.~(\ref{eq:w}), integrating, and using
Eq.~(\ref{eq:K}) yields,
\begin{eqnarray}
\frac{d\tilde{w}_0}{dt}(t) &=& 2 \frac{d}{dt} \coth K(t) \nonumber \\
&=& 2 \frac{d}{dt} \cosh \frac{\tilde{\zeta}_0(t)}{2}  \label{eq:dw0dt} \\
&=& - \frac{\tilde{q}_0(t)}{2} \,  \frac{d\tilde{\zeta}_0(t)}{dt} \nonumber
\end{eqnarray}
where we have used the identity, $\cosh 2z = -\coth\log\tanh z$.
Recall that the leading-order zeta potential, $\tilde{\zeta}_0(t)$, is
related via Eq.~(\ref{eq:phid}) to the leading-order bulk current
density, $\bar{j}_0(t)$, or interfacial voltage, $\tilde{\Psi}_0(t)$,
given by Eqs.~(\ref{eq:jexact}) and (\ref{eq:F}). Integrating
Eq.~(\ref{eq:dw0dt}) and requiring $\tilde{w}_0=0$ for
$\tilde{\zeta}_0=0$, we also obtain a simple expression for the excess
concentration,
\begin{equation}
\tilde{w}_0 = 4\, \sinh^2 \frac{\tilde{\zeta}_0}{4} ,
\label{eq:w0}
\end{equation}
which also holds for the static Gouy-Chapman solution.  Of course,
this is another sign that (at leading order) a thin double layer stays
in quasi-equilibrium, even while charging.

\subsection{ The Sign of the Donnan Effect }

Before proceeding to calculate the bulk dynamics, we comment on the
sign of the excess concentration in the diffuse part of the double
layer, which corresponds to a {\it positive} adsorption of neutral
salt. In contrast, it is commonly believed that double-layer salt
adsorption is always negative, resulting in an excess neutral
concentration in the nearby bulk electrolyte. Lyklema calls this the
``Donnan effect'' with reference to Donnan's general papers on
membrane equilibria~\cite{donnan11a,donnan11b,donnan24} and describes
how it is used to infer surface areas from experimental measurements
of concentration variations upon charging (``salt
sieving")~\cite{lyklema,lyklema1}. In the present case of
diffuse-layer adsorption, the following argument is given: Since the
equilibrium co-ion concentration in the double layer near a charged
surface is reduced compared to the bulk, there must be an excess of
co-ions, and hence an excess of neutral concentration (and
counter-ions) in the nearby bulk.

How can this belief be reconciled with our analytical and numerical
results, which clearly demonstrate the opposite effect in a model
problem? The difference is that we consider a {\it finite} system with
global ion conservation, while Lyklema considers an open system into
which ions are apparently injected. We also explicitly calculate the
time-dependent response to interfacial charging, while he describes the
steady state after new ions somehow arrive ``from infinity''.

Our conclusion is therefore the opposite: Since the equilibrium
counter-ion concentration is enhanced in the double layer, there must
be a depletion of counter-ions, and hence a reduction in neutral
concentration (and co-ions) in the nearby bulk. Gouy's nonlinear
theory shows that at large voltages the excess of counter ions exceeds
the reduction in co-ions in the diffuse layer, so this result is quite
intuitive.

In real systems, it may be that compact-layer effects, such as the
specific adsorption of ions on the surface, can lead to overall
negative adsorption by the double layer. According to the
Nernst-Planck-Poisson equations, however, the average concentration of
all ions (regardless of species) is always increased in the diffuse
part of the double layer relative to the bulk in any finite system.
The associated local depletion of neutral salt has important
implications for time-dependent electrokinetic phenomena at
polarizable surfaces (section ~\ref{sec:colloids}) since bulk
concentration gradients alter electric fields and produce
diffusio-osmotic slip.

\subsection{ Bulk Diffusion at Two Time Scales }

We now proceed to calculate how the bulk concentration is
depleted in time and space during double-layer charging in our model
problem.  The matching condition, Eq.~(\ref{eq:wmatch}), seems to contradict
the analysis above, since it introduces a new time variable,
$\bar{t}$. However, this is the same time scale for the first-order
(diffusive) dynamics in the bulk,
\begin{equation}
\frac{\partial \bar{c}_1}{\partial \bar{t}}
= \frac{1}{\epsilon} \,  \frac{\partial \bar{c}_1}{\partial t}
= \frac{\partial^2 \bar{c}_1}{\partial x^2} .   \label{eq:c1eq}
\end{equation}
We must solve this equation starting from $\bar{c}_1(x,0)=0$ with a
time-dependent prescribed flux at $x=-1$ given by
Eqs.~(\ref{eq:wmatch}) and (\ref{eq:dw0dt}). We also enforce symmetry
about the origin, Eq.~(\ref{eq:sym}). 

The Laplace transform of the solution is:
\begin{eqnarray}
\hat{\bar{c}}_1(x,s)
&=& - \frac{\sqrt{s}\,
    \cosh(x\sqrt{s})}{\sinh(\sqrt{s})}\, \int_0^\infty   
    e^{-s\bar{t}} \tilde{w}_0(\bar{t}/\epsilon)
    d\bar{t}   \nonumber \\
&=& \int_0^\infty e^{-s\bar{t}} \bar{c}_1(x,\bar{t}) d\bar{t}   
    \label{eq:c1lap} 
\end{eqnarray}
where $\tilde{w}_0(t)$ is determined by $\tilde{\zeta}_0(t)$ from
Eq.~(\ref{eq:w0}). The prefactor,
\begin{equation}
\hat{G}(s) = \frac{\cosh(x\sqrt{s})}
	{\sqrt{s}\, \sinh(\sqrt{s})} ,  \label{eq:c1t}
\end{equation}
is the Laplace transform of $G(\bar{t})$, the Green function for
the diffusion equation, Eq. ~(\ref{eq:c1eq}), for a sudden unit flux
of ions at time $\bar{t} = 0^+$ injected at the boundary:
\begin{equation}
G(x,0)=0, \ \ \frac{\partial G}{\partial x}(-1,\bar{t}) =
\delta^+(\bar{t}) .  \label{eq:c1bc1}
\end{equation}
The same Green function also arises in the equivalent problem of an
initial unit source adjacent to a reflecting wall,
\begin{equation}
G(x,0) = \delta(x+1^+), \ \ \frac{\partial G}{\partial
x}(-1,\bar{t}) = 0 . \label{eq:c1bc2}
\end{equation}
In this form, the Green function can be obtained by inspection,
\begin{equation}
G(x,\bar{t}) = \frac{1}{\sqrt{\pi
\bar{t}}} \, \sum_{m=-\infty}^\infty e^{-(x - 2m+1)^2/4\bar{t}} ,
\label{eq:c1profile}
\end{equation}
using the method of images.

Since $\hat{\bar{c}}_1(x,s)$ is expressed as a product of two
Laplace transforms, Eq.~(\ref{eq:c1lap}), the inverse is equal to 
the convolution of the two original functions:
\begin{equation}
\bar{c}_1(x,\bar{t}) = -\int_0^{\bar{t}} d\bar{t}^\prime \, 
G(x,\bar{t}^\prime-\bar{t}) \, \frac{\partial \tilde{w}_0}{\partial \bar{t}}
\left(\frac{\bar{t}^\prime}{\epsilon}\right). \label{eq:c1int1} 
\end{equation}
This form clearly demonstrates that the boundary forcing occurs over the
fast, charging time, $t = \bar{t}/\epsilon$, while the response
described by the Green-function kernel occurs over the slow, diffusion
time, $\bar{t}$. The separation of time scales is apparent in
the equivalent expression,
\begin{equation}
\bar{c}_1(x,t) = - \int_0^t dt^\prime
G\left(x,\epsilon(t^\prime-t)\right)\,
\tanh\left(\frac{\tilde{\zeta}_0(t^\prime)}{2}\right)\,
\bar{j}_0(t^\prime) \label{eq:c1j}
\end{equation}
which can be derived from Eq.~(\ref{eq:c1int1}) using
Eq.~(\ref{eq:dw0dt}). This form shows explicitly how solution for the
current at leading order, Eq.~(\ref{eq:jexact}), fully determines the
bulk concentration at first order.

\begin{figure}
\includegraphics[width=\linewidth]{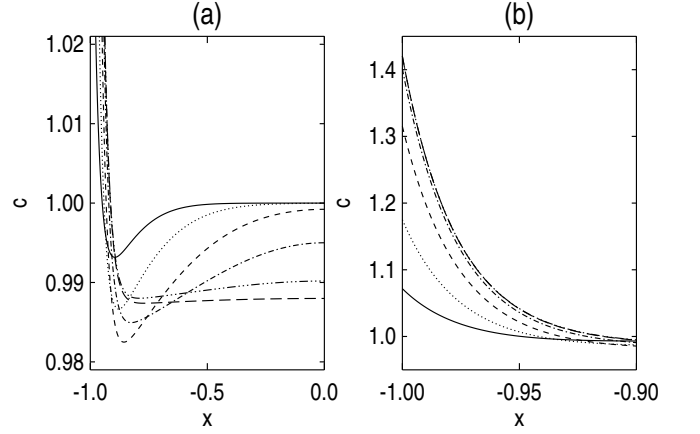}
\caption{ Weakly nonlinear dynamics for $v=1$, $\epsilon=0.05$,
$\delta=0.1$, showing the effect of bulk diffusion. The concentration
from the full numerical solution is shown in the half cell (a) and in
the diffuse layer (b) for $t=0.5$ (solid), $t=1$ (dot), $2$ (dash),
$4$ (dot-dash), $8$ (dot-dot-dot-dash), and $20$ (long
dash). \label{fig:weak} }
\end{figure}

Before further analysis of the exact solution for $c_1(x,t)$, we
describe its physical significance. The bulk concentration at
first-order exhibits diffusive relaxation at two different time
scales, $t=O(1)$ and $\bar{t}=O(1)$, or with units, $\tau =
O(\lambda_DL/D)$ and $\tau = O(L^2/D)$, respectively.  For $t=O(1)$
and $\bar{t} = O(\epsilon)$, the initial double-layer charging process
proceeds without any significant changes in concentration at the bulk
length scale, $x = O(1)$. During this phase, each diffuse-charge layer
acquires an $O(\epsilon)$ amount of excess concentration, given by
Eq.~(\ref{eq:w0v}).  This excess concentration has been acquired by a
diffusive process, which at this time scale corresponds to a bulk
diffusion layer of $O(\sqrt{\epsilon})$ width near each electrode.
This implies an $O(\sqrt{\epsilon})$ depletion of the neutral salt
concentration in the bulk diffusion layers. These scaling arguments
are confirmed by Fig.~\ref{fig:compare}(c) for $v=1$ and
$\epsilon=0.05$, where at time $t = 1$ (or $\bar{t} = \epsilon$) the
diffusion layers are roughly of width $\sqrt{2\bar{t}} =
\sqrt{2\epsilon} \approx 0.3$. The formation and spreading of the
diffusion layers is also shown in more detail in Fig.~\ref{fig:weak}.

\subsection{ Evolution of the Diffusion Layers }

In the previous section, we derived the time-dependent outer
approximation, Eq.~(\ref{eq:cout}), to first order,
\begin{equation}
\bar{c}(x,t) \sim 1 + \epsilon \, \bar{c}_1(x,t),  \label{eq:cc1}
\end{equation}
which displays dynamics at both the RC time and the bulk diffusion
time. The result, Eqs.~(\ref{eq:c1profile})--(\ref{eq:c1j}), is fairly
complicated, so in this section we try to gain some simple analytical
insight. In the limit $\epsilon \rightarrow 0$, the initial charging
process at the time scale $\bar{t} = O(\epsilon)$ is instantaneous,
and we are left with only the slow relaxation of the bulk diffusion
layers. Explicitly taking this limit in Eq.~(\ref{eq:c1lap}) with
$\bar{t} = \epsilon t$ fixed, 
\begin{equation}
%\lim_{\epsilon\rightarrow 0} \hat{\bar{c}}_1(x,s) 
%&=& -\tilde{w}_0(\infty) \hat{G}(s) \nonumber \\ 
\lim_{\epsilon\rightarrow 0}
\bar{c}_1(x,\bar{t}) = -\tilde{w}_0(\infty) G(x,\bar{t}) ,   \label{eq:c1approx}
\end{equation}
we see that the slow-scale evolution of the diffusion layers is 
given by the Green function, $G(x,\bar{t})$, with a source of
strength, $-\tilde{w}_0(\infty)$, equal to the leading-order total salt
adsorption. According to Eqs.(\ref{eq:phid}) and (\ref{eq:w0}) with
$\bar{j}_0(\infty)=0$, this is given by
\begin{equation}
\tilde{w}_0(\infty) = 4\, \sinh^2 \frac{f^{-1}(v)}{4}, 
\label{eq:w0v}
\end{equation}
where
\begin{equation}
f(\zeta) = \zeta + 2\delta \sinh(\zeta/2) \label{eq:fzeta}
\end{equation}
which reduces to 
\begin{equation}
\tilde{w}_0(\infty) = 4\, \sinh^2 \frac{v}{4},  \label{eq:wtd0}
\end{equation}
in the absence of any compact layers ($\delta=0$).

\begin{figure}
\includegraphics[width=\linewidth]{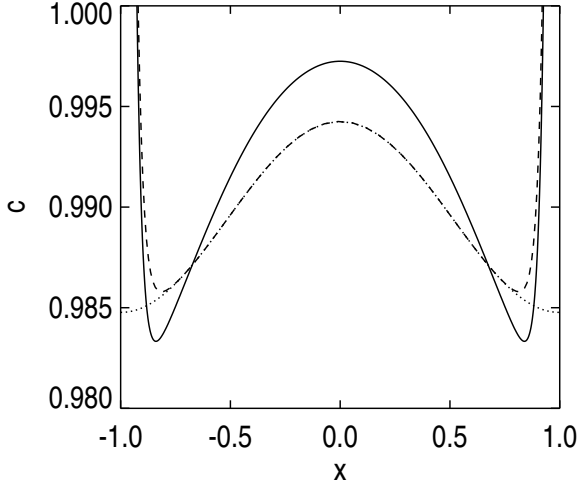}
\caption{ \label{fig:c1} Simple approximations of the bulk diffusion
layers for weakly nonlinear charging dynamics with $v=1$,
$\epsilon=0.05$, $\delta=0.1$. The full numerical solution (solid) is
compared with the approximate first-order expansion at the diffusion
time scale, given by Eqs.~(\ref{eq:c1approx}), (\ref{eq:c1profile}),
and (\ref{eq:cc1}). Also, shown is the latter plus the zeroth-order
inner approximation, Eq.~(\ref{eq:cin0}), for the diffuse layers
(dashed). }
\end{figure}

This simple approximation describes two diffusion layers created at
the electrodes slowly invading the entire cell. At first, they have
simple Gaussian profiles,
\begin{equation}
\bar{c}(x,\bar{t}) \sim 1 - \frac{\epsilon
\tilde{w}_0(\infty)}{\sqrt{\pi \bar{t}}} \left[ e^{-(x +
1)^2/4\bar{t}} + e^{-(x -1)^2/4\bar{t}} \right]  \label{eq:cinit}
\end{equation}
for $\bar{t} \ll 1$, which is qualitatively consistent with the
numerical results in Fig.~\ref{fig:compare}(c). To attempt a
quantitative comparison, we also need $t \gg 1$ to use
Eqs.~(\ref{eq:c1approx}) and (\ref{eq:c1profile}). As shown in
Fig.~(\ref{fig:c1}), the approximation is reasonable for $t=3$ with an
error of roughly $\epsilon^2 = 0.0025$. The two diffusion layers
eventually collide, and the concentration slowly approaches a
(reduced) constant value,
\begin{equation}
\bar{c}(x,\bar{t}) \sim 1 - \epsilon \tilde{w}_0(\infty)   \label{eq:cinf}
\end{equation}
for $\bar{t} \gg 1$, as expected from the steady-state excess
concentration in the double layers. (This result may be checked by
replacing the sum in Eq.~(\ref{eq:c1profile}) with an integral in the
limit $\bar{t} \rightarrow \infty$.)

\subsection{ Bulk Concentration Polarization }

As mentioned above, the bulk charge density remains very small,
$\bar{\rho} = O(\epsilon^3)$, even during double-layer relaxation, but
changes in neutral bulk concentration affect the potential at first
order. Substituting the outer expansions into Eq.~(\ref{eq:rho}) and
collecting terms at $O(\epsilon)$, we have
\begin{equation}
0 = \frac{\partial}{\partial x}\left[ \bar{c}_0 
  \frac{\partial\bar{\phi}_1}{\partial x} + \bar{c}_1 
  \frac{\partial\bar{\phi}_0}{\partial x} \right].
\end{equation}
This is easily integrated using $\bar{c}_0 = 1$ to obtain the
first-order contribution to the bulk electric field,
\begin{equation}
\frac{\partial\bar{\phi}_1}{\partial x} = \bar{j}_1(t) - \bar{j}_0(t)
\, \bar{c}_1(x,t),
\end{equation}
where the second term describes concentration polarization, i.e. the
departure from a harmonic potential, which would be predicted by Ohm's
law.  The first term is a uniform bulk field (or current) determined
by first-order perturbation in double-layer charge. This follows from
the matching condition, Eq.~(\ref{eq:qmatch}), at first order:
\begin{equation}
\frac{d\tilde{q}_1}{dt} = \bar{j}_1(t).
\end{equation}
where $\tilde{q}_1(t)$ is obtained by solving the inner problem at
first order.

\subsection{ Perturbations in Double-layer Structure }

Unfortunately, the first-order inner problem is difficult to solve
analytically because the perturbed concentration profiles are no
longer in thermal equilibrium during the initial charging phase. To
see this, note that the time derivatives in Eqs.~(\ref{eq:cin}) and
(\ref{eq:rhoin}) contribute nonzero (but known) terms at first order,
\begin{eqnarray}
\frac{\partial \tilde{c}_0}{\partial t} &=& 
\frac{\partial}{\partial y}\left( \frac{\partial \tilde{c}_1}{\partial y} +
\tilde{\rho}_0 \frac{\partial \tilde{\phi}_1}{\partial y} +
\tilde{\rho}_1 \frac{\partial \tilde{\phi}_0}{\partial y}\right)
\label{eq:c1in} \\
\frac{\partial \tilde{\rho}_0}{\partial t} &=& 
\frac{\partial}{\partial y}\left( \frac{\partial \tilde{\rho}_1}{\partial y} +
\tilde{c}_0 \frac{\partial \tilde{\phi}_1}{\partial y} +
\tilde{c}_1 \frac{\partial \tilde{\phi}_0}{\partial y}\right)
\label{eq:rho1in} \\
 -\frac{\partial^2 \tilde{\phi}_1}{\partial y^2} &= &\tilde{\rho}_1
\label{eq:phi1in} 
\end{eqnarray}
although one still solves a system of linear ordinary differential
equations in $y$ at each $t$, since $\tilde{c}_0(y,t)$,
$\tilde{\rho}_0(y,t)$, and $\tilde{\phi}_0(y,t)$ are known.

The general problem seems daunting, but some progress can
be made at the scale of bulk diffusion, $\bar{t} = O(1)$ or $t =
O(\epsilon^{-1})$, where the leading-order concentration profiles
remain in thermal equilibrium, without any explicit time
dependence. This will give us some insight into secondary charge
relaxation at the time scale of bulk diffusion. In this limit, the
Equations~(\ref{eq:c1in}) and (\ref{eq:rho1in}) can be integrated to
obtain: \begin{eqnarray} - \frac{\partial \tilde{c}_1}{\partial y}
&\sim& \tilde{\rho}_0 (y,\infty)\frac{\partial
\tilde{\phi}_1}{\partial y} + \tilde{\rho}_1 \frac{\partial
\tilde{\phi}_0}{\partial
 y}(y,\infty) 
\label{eq:c1inb} \\
- \frac{\partial \tilde{\rho}_1}{\partial y} &\sim&
\tilde{c}_0 (y,\infty)\frac{\partial \tilde{\phi}_1}{\partial y} +
\tilde{c}_1 \frac{\partial \tilde{\phi}_0}{\partial
 y}(y,\infty) 
\label{eq:rho1inb} 
\end{eqnarray}
after applying the usual van Dyke matching conditions.
Substituting from Poisson's equation, $\tilde{\rho}_n = -\partial^2
\tilde{\phi}_n/\partial y^2$, at orders $n=0,1$ into
Eq.~(\ref{eq:c1inb}), integrating, and applying matching again, we
obtain:
\begin{equation}
\tilde{c}_1(y,\bar{t}) \sim \frac{\partial \tilde{\phi}_0}{\partial
 y}(y,\infty) \,  \frac{\partial \tilde{\phi}_1}{\partial
 y}(y,\bar{t}) + \bar{c}_1(-1,\bar{t})
\end{equation}
for $\bar{t} > 0$ and $t = \bar{t}/\epsilon \gg 1$.  From the previous
section, we also have the leading-order inner concentration,
\begin{equation}
\tilde{c}_0(y,\infty) = \bar{c}_0(-1,\bar{t}) + \frac{1}{2}
\tilde{E}_0(y,\infty),
\end{equation}
where $\bar{c}_0(-1,\bar{t})=1$ is the leading-order outer
concentration, and 
\begin{equation}
\tilde{E}_0(y,\infty) = -\frac{\partial \tilde{\phi}_0}{\partial
y}(y,\infty) = 2 \sinh \frac{\tilde{\psi}_0(y,\infty)}{2},
\end{equation}
is the leading-order inner electric field in steady-state.  Finally,
we substitute these expressions into Eq.~(\ref{eq:rho1inb}) and use
Eq.~(\ref{eq:rho1in}) to obtain a master equation for the first-order
inner electric field, $\tilde{E}_1(y,\bar{t})= -\frac{\partial
\tilde{\phi}_1}{\partial y}(y,\bar{t})$, at the bulk-diffusion time
scale:
\begin{equation}
\frac{\partial^2 \tilde{E}_1}{\partial y^2}= \left(1 + \frac{3}{2}
\tilde{E}_0^2\right) \tilde{E}_1 +  \bar{c}_1 \tilde{E}_0
\label{eq:E1master} .
\end{equation}
This linear equation with a non-constant coefficient must be solved
subject to the boundary conditions, $\tilde{E}_1(\infty,\bar{t}) = 0$
and $\tilde{E}_1(0,\bar{t}) = -\tilde{q}_1(\bar{t})$. The perturbation
of the total charge, $\tilde{q}_1(\bar{t})$ is obtained by another
integration of the field to get the first-order inner potential, while
applying the Stern boundary condition.

For our purposes here, it suffices to point out that the spatial
profile of the first-order inner electric field in
Eq.~(\ref{eq:E1master}) varies with the outer concentration,
$\bar{c}_1(-1,\bar{t})$, at the slow time scale of bulk
diffusion. Notably, this can lead to a secondary relaxation of the
total diffuse charge, in response to the evolution of the diffusion
layers. We observe this slow relaxation phase in our numerical
solutions of the full equations, especially at large voltages. In
particular, it is presumably associated with the non-monotonic
charging profile for $v = 4$ shown in Fig.~\ref{fig:q}(c). A detailed
analysis of this interesting effect from the setup above would require
solving the first-order inner problem numerically, so we leave it for
future work.

\section {Strongly Nonlinear Dynamics }
\label{sec:strong}

\subsection{ Steady State and the Dukhin Number }

We stress again that the asymptotic expansions derived above are valid
in the limit of thin double layers, $\epsilon \rightarrow 0$, with the
other two dimensionless parameters, $v$ (applied voltage) and $\delta$
(relative compact-layer capacitance), held constant. For any fixed
$\epsilon > 0$, there is no guarantee that the approximation remains
accurate as the other parameters are varied. Having just calculated
the bulk concentration to first order in the regular
expansion, Eq.~(\ref{eq:cout}), we can now check {\it a posteriori}
under what conditions it remains a good approximation.

A simple check involves the constant bulk concentration,
Eq.~(\ref{eq:cinf}), after the charging process is completed.  The
assumption that the first correction is much smaller than the leading
term requires, $\alpha_s \equiv \epsilon \, \tilde{w}(\infty) \ll 1$.
Linearizing Eq.~(\ref{eq:fzeta}) for $\delta \ll 1$, we can write this
condition in a closed form:
\begin{equation}
4 \, \epsilon \, \sinh^2 \left( \frac{v}{4(1+\delta)} \right) \ll 1
\end{equation}
Putting the units back, we have  
\begin{equation}
\alpha_s(\zeta_0) = \frac{4\lambda_D}{L} \, \sinh^2 \left( \frac{ze
\zeta_0}{4 kT} \right) \ll 1 \label{eq:valid-steady}
\end{equation}
where $\zeta_0 \approx V/(1+\delta)$ is the steady-state zeta
potential, long after the DC voltage is applied. 

The condition, $\alpha_s(\zeta_0) \ll 1$, for the validity of the {\it
steady-state} asymptotic expansion is identical to that of small
Dukhin number, $\Du(\zeta_0) \ll 1$, from Eq.~(\ref{eq:Du}) in the
limit of no electro-osmosis ($m=0$), which may seem surprising since
there is no surface conduction in our one-dimensional model
problem. (Hence, we use the symbol $\alpha_s$ rather than $\Du$.)  The
reason is that in both cases --- Dukhin's problem of electrophoresis
of highly charged particles in weak applied fields and ours of electrode
screening in strong applied fields --- the double layer absorbs a significant
amount of neutral salt from the bulk (reverse Donnan effect). 

Net charge adsorption relative to the point of zero charge is measured by the
{\it total} zeta potential,
\begin{equation}
\zeta_{tot} = \zeta_{eq} + \zeta_{ind}
\end{equation}
where $\zeta_{eq}$ is the uniform equilibrium zeta potential
(reflecting the initial surface charge) and $\zeta_{ind}$ is the
non-uniform induced zeta potential (resulting from diffuse-charge
dynamics). In Dukhin's problem, the former may be large,
$\Du(\zeta_{eq}) > 1$, but the latter is always small, $\zeta_{ind}
\ll kT/ze$, so that the charging dynamics is linearized (or
ignored). In our model problem, the situation is reversed: We assume
$\zeta_{eq}=0$ (for simplicity), but we allow for a large applied
voltage, $\zeta_{ind} \approx v/(1+\delta) > kT/ze$, in which case the
dynamics is nonlinear. In both cases, the steady state is well
described by weakly nonlinear asymptotics as long as
$\alpha_s(\zeta_{tot}) = \Du(\zeta_{tot}) \ll 1$. When this condition is
violated, double-layer charging and surface conduction may cause
significant changes in the steady-state bulk concentration.

\subsection{ Breakdown of Weakly Nonlinear Asymptotics }

In general, weakly nonlinear {\it dynamics} breaks down at
somewhat smaller voltages, where $\zeta_{tot} > kT/e$ but
$\alpha_s(\zeta_{tot}) = \Du(\zeta_{tot}) \ll 1$, because neutral-salt
adsorption causes a {\it temporary, local depletion} of bulk
concentration exceeding that of the steady state, after diffusional
relaxation. In our model problem, the maximum change in bulk
concentration occurs just outside the diffuse layers at $x = \pm 1$,
just after the initial charging process finishes at time scale, $t =
1$ or $\bar{t} = \epsilon$. From 
Eq.~(\ref{eq:cinit}), we have the first two terms of the weakly
nonlinear asymptotic expansion there:
\begin{equation}
\bar{c}(\pm 1,\epsilon) \sim 1 - \sqrt{\frac{\epsilon}{\pi}} \,
  \tilde{w}_0(\infty).
\end{equation}
At that time, the newly created diffusion layers have spread to
$O(\sqrt{\epsilon})$ width, so the concentration is depleted locally
by $O(\epsilon/\sqrt{\epsilon})=O(\sqrt{\epsilon})$, which is much
more than the uniform $O(\epsilon)$ depletion remaining after bulk
diffusion.
 
Therefore, in order for the time-dependent correction term to be
uniformly smaller than the leading term, we need
\begin{equation}
\alpha_d \equiv  \sqrt{\frac{\epsilon}{\pi}} \, \tilde{w}_0(\infty) =
\frac{\alpha_s}{\sqrt{\pi \epsilon}} \ll 1 .
\end{equation}
The relevant dimensionless parameter,  
\begin{equation}
\alpha_d(\zeta_{tot}) = 4 \sqrt{\frac{\lambda_D}{\pi L}} \, \sinh^2 \left( \frac{ze
\zeta_{tot}}{4 kT} \right)  
\end{equation}
is larger than $\alpha_s(\zeta_{tot})$ (and the Dukhin number) by a factor
of $\sqrt{L/\pi\lambda_D}$ in the limit of thin double layers.  For
weakly nonlinear dynamics to hold, the applied voltage
cannot greatly exceed the thermal voltage,
\begin{equation}
\zeta_{tot} \approx \frac{V}{1+\delta} < \frac{kT}{ze} \, \log \frac{
  L}{\lambda_D} , \label{eq:valid-dyn}
\end{equation}
even for very thin double layers, $\lambda_D \ll L$, due to the
logarithm.  In comparison, the applied voltage can be twice as large
before the steady-state bulk concentration is significantly affected
(and surface conduction becomes important in higher dimensions).
This
could have interesting consequences for induced-charge
electrokinetic phenomena~\cite{bazant03} at moderate applied voltages
where $\alpha_d > 1$ but $\alpha_s = \Du < 1$.

\subsection{ Strongly Nonlinear Asymptotics }

\begin{figure*}
\includegraphics[width=6in]{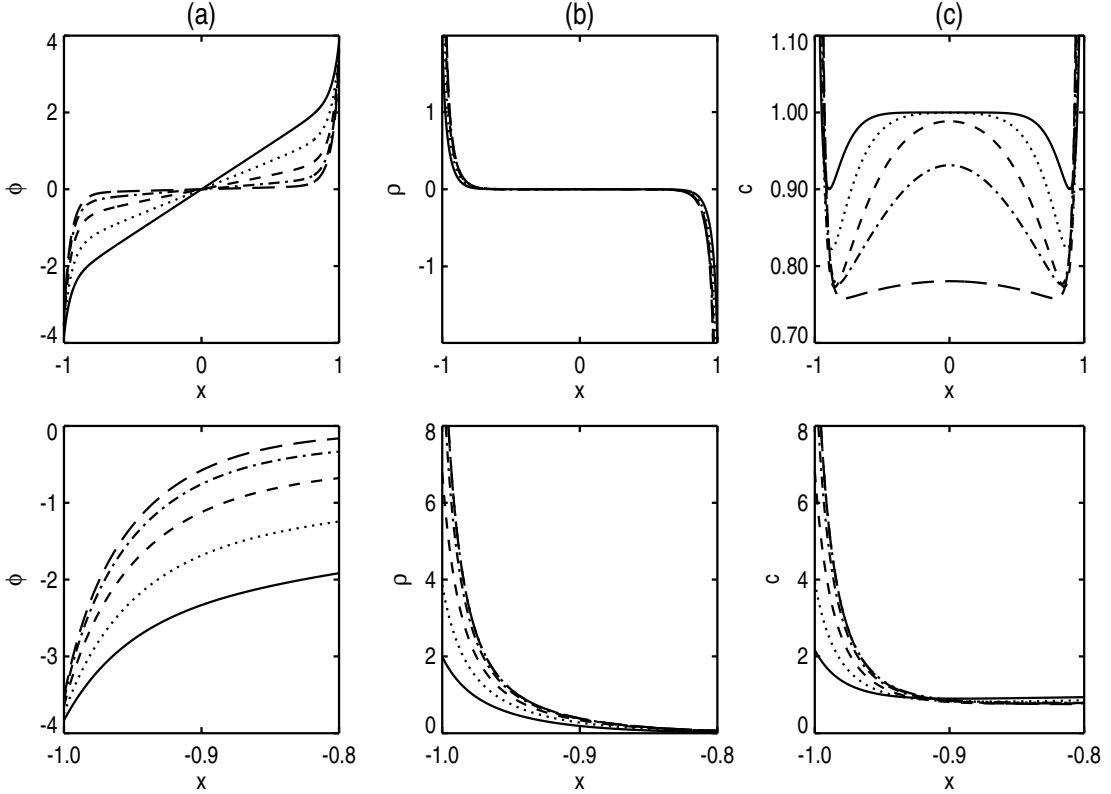}
\caption{ Strongly nonlinear charging dynamics for $v=4$ with
$\epsilon=0.05$ and $\delta=0.1$. The potential (a), charge density
(b), and concentration are shown in the half cell (top) and in the
diffuse layer (bottom) for $t=0.5$ (solid), $1$ (dot), $2$ (dash), $4$
(dot-dash), $8$ (dot-dot-dot-dash), and $20$ (long
dash). \label{fig:strong} }
\end{figure*}

When condition (\ref{eq:valid-dyn}) is violated, electrochemical
relaxation becomes much more complicated because double-layer charging
is coupled to bulk diffusion. As long as $\alpha_d<1$, 
however, the
bulk remains quasi-neutral at all times. This regime of strongly
nonlinear dynamics is demonstrated by the numerical solution in
Fig.~\ref{fig:strong}, for $v = 4$, $\epsilon=0.05$, and $\delta=0.1$,
in which case $\alpha_d = 0.545$.
%
% gnuplot> print 4*sqrt(0.05/pi)*(sinh(1.0/1.1)**2)
% 0.545371019206225
%
In spite of the substantial $O(1)$
amount of charge transfered from one diffuse layer to the other, each
retains almost exactly the same $O(\epsilon)$ width as at lower 
voltages, and bulk electro-neutrality remains an excellent
approximation for all times. The initial charging process up
to $t \approx 1$ creates a diffusion layer of neutral salt which
relaxes into the bulk at the scale $\bar{t} \approx 1$ (or $t =
\bar{t}/\epsilon \approx 20$).

In the strongly nonlinear regime, if $\alpha_s$ is not too small,
double-layer charging is slowed down so much by nonlinearity that
it continues to occur as the bulk diffusion layers evolve. One way to
see this is that the effective RC time for the late stages of
charging in Eq.(\ref{eq:jlate}) is
\begin{equation}
t_c(v) = C_i(v) \approx \cosh \frac{v}{2} \approx 2 \sinh^2 \frac{v}{4}
\approx \frac{\alpha_s}{2\epsilon}  
\end{equation}
where we use the leading-order approximation of the differential
capacitance, Eq. (\ref{eq:Ca}), for $\delta \ll 1$. In units of the
bulk diffusion time, the nonlinear relaxation time is $\bar{t}_c =
\epsilon t_c = \alpha_s/2$. 

To make analytical progress, one would consider the joint limits
\begin{equation}
\epsilon \rightarrow 0 \ \ \mbox{ and } \ \ v \rightarrow \infty \ \
\mbox{with } \ \ \alpha_d(v)>0 \mbox{ fixed}
\end{equation}
and expect the approximations to remain acceptable at somewhat
larger voltages, as long $\alpha_s(v) < 1$.  Such analysis is beyond
the scope of this article, but at least we indicate how the leading
order approximation would be calculated. (Going beyond leading order
seems highly nontrivial.) 

At leading order in the bulk, we have the
usual equations for a neutral binary electrolyte (with equal ionic diffusivites),
\begin{equation}
\frac{\partial \bar{c}_0}{\partial \bar{t}} = \frac{\partial^2
\bar{c}_0}{\partial x^2} \ \ \ \mbox{and} \ \ \ \frac{\partial
}{\partial x} \left(\bar{c}_0 \, \frac{\partial
\bar{\phi}_0}{\partial x}\right) = 0
\end{equation}
with $\bar{\rho} = O(\epsilon^2)$. Integrating the second equation, we
obtain a constant, uniform current density, $\bar{j}_0(t)$,  as before,
but the electric field is modified by concentration polarization,
\begin{equation}
\frac{\partial \bar{\phi}_0}{\partial x} =
\frac{\bar{j}_0(t)}{\bar{c}_0(x,\bar{t})}.
\end{equation}
The effective boundary conditions come from asymptotic matching with
the diffuse layers as before, 
\begin{equation}
\epsilon \, \frac{d\tilde{q}_0}{d\bar{t}} = \bar{j}_0(t) \ \ \ \mbox{and
} \ \ \ \epsilon \, \frac{d\tilde{w}_0}{d\bar{t}} =  \frac{\partial
\bar{c}_0}{\partial x}(-1,\bar{t})
\end{equation}
only now the diffusive flux entering the diffuse layers (second equation)
appears at leading order.  The ionic concentrations retain 
Gouy-Chapman equilibrium profiles modified
quasi-statically by the evolving nearby bulk concentration:
\begin{eqnarray}
\tilde{q}_0(\bar{t}) &=& - 2 \sqrt{\bar{c}_0(-1,\bar{t})}
\sinh\left(\frac{\tilde{\zeta}_0(\bar{t})}{2}\right)   \\
\tilde{w}_0(\bar{t}) &=& 4 \sqrt{\bar{c}_0(-1,\bar{t})}
\sinh^2\left(\frac{\tilde{\zeta}_0(\bar{t})}{4}\right)  
\end{eqnarray}
where
\begin{equation} 
\tilde{\zeta}_0(\bar{t}) - \tilde{q}_0(\bar{t})\,  \delta =
\tilde{\Psi}_0(\bar{t})  = - v -
\bar{\psi}_0(-1,\bar{t}) .
\end{equation}
It seems exact solutions are not possible in terms of elementary
functions.
The equations are ``stiff'', since they involve a short time
scale, $\bar{t} = \epsilon$, for the initial phases of charging, but
at least the spatial boundary layers have been ``integrated out'',
which is convenient for numerical solutions.

\subsection{ Space Charge at Very Large Voltages }

We close this section by noting some intriguing, new possibilities, 
further into the strongly nonlinear regime.  At large voltages, such
that $\alpha_d > 1$, it seems a {\it transient space charge layer}
should form since the bulk concentration would be depleted almost
completely near the diffuse layers by the initial charging process.
In steady-state problems of Faradaic conduction, it is well known that
double-layer structure is altered from its Gouy-Chapman equilibrium
profile at a limiting current~\cite{smyrl67} and may turn into an
extended space charge layer above a limiting
current~\cite{rubinstein79}, but here we see that similar effects may
also occur temporarily with large time-dependent voltages, in the
absence of any Faradaic processes (at blocking electrodes).  At still
larger voltages, such that $\alpha_s > 1$, double layer charging
consumes most of the bulk concentration, presumably leaving the entire
bulk region in a state of ``space charge''.  

Such situations may seem quite exotic in macroscopic systems, where
$\epsilon = \lambda_D/L$ is extremely small, but in microsystems
perhaps they could occur. The mathematical model neglects bulk
reactions (e.g. leading to hydrogen bubble formation), nonlinear
dielectric properties, electro-convection, or other effects which may
hinder the formation of space charge in real systems. Nevertheless,
the rich nonlinear behavior of the model merits further mathematical
study, as a challenging problem in time-dependent boundary-layer
theory.

\section{ Beyond the Model Problem }
\label{sec:concl}

We conclude by discussing more general situations, which contain some
new physics, absent in our simple model problem.  For thin double layers,
the same methods of asymptotic analysis could be applied to derive
effective equations in which the double layers are incorporated into
boundary conditions, better suited for analytical or numerical
work. Here, we simply sketch the results and suggest some other model
problems for further study. 

\subsection{ Two or More Dimensions }
 
In the weakly nonlinear regime, where $\alpha_d < 1$ for all times
over all double layers, our analysis  extends trivially to higher
dimensions, as long as the surface curvature does not introduce another
length scale much
smaller than $L$. In that case, the double layers are
locally ``flat'', and the boundary-layer calculations remain
unchanged. Following the same procedure, we find that the bulk concentration is
uniform at leading order, $\bar{c}_0 = 1$, and the bulk potential,
$\bar{\phi}(\rb,t)$, is a harmonic function,
\begin{equation}
\nabla^2 \bar{\phi}_0 = 0 ,  \label{eq:laplace}
\end{equation}
subject to a (dimensionless)  RC boundary condition at each
electrode surface,
\begin{equation}
\frac{\partial \tilde{q}_0}{\partial t} = \tilde{C}(\bar{\phi}_0 -
\phi_e) \, \frac{\partial (\bar{\phi}_0-\phi_e)}{\partial t} =
\nb\cdot\del \bar{\phi}_0 ,   \label{eq:RCBC}
\end{equation}
where $\nb$ is the unit normal pointing into the electrolyte and  
$\phi_e(\rb,t)$ is the local electrode potential relative to the
solution. The latter is equal to the local applied voltage plus the
equilibrium zeta potential:
\begin{equation}
\phi_e(\rb,t) = V(\rb,t) + \zeta_{eq}(\rb)
\end{equation}
which accounts for any pre-existing double-layer charge (neglected in
our calculations above).  A Neumann boundary condition,
$\nb\cdot\del\bar{\phi}_0$, is imposed at any inert, non-polarizable
%%%% QQQ Missing right hand side in the above boundary condition????
% NO. The prior eq is the boundary condition (it is an ODE).
surface, such as a channel side wall. 

Another complication in two or more dimensions is the possibility of
electro-osmotic flow. The fluid velocity in the bulk usually satisfies
the Stokes equations, which may be unsteady for high-frequency
forcing. In the weakly nonlinear regime, the classical
Helmholtz-Smoluchowski formula gives the fluid slip in terms of the
local zeta potential and tangential bulk electric
field~\cite{hunter,russel,lyklema}.

Equations ~(\ref{eq:laplace}) and (\ref{eq:RCBC}) model the
electrolyte as a bulk Ohmic resistor with a capacitor skin at
electrode interfaces.  The linearized version of these equations (with
$\tilde{C}$ = constant) has been studied extensively, e.g.\ in the
context of metallic colloids~\cite{murtsovkin96,simonov77}, AC
electro-osmosis~\cite{ramos99,green00a,gonzalez00}, AC
pumping~\cite{ajdari00}, and other phenomena of induced-charge
electro-osmosis~\cite{bazant03,squires03}. The nonlinear version,
however, has apparently not been analyzed, even though it may have
relevance for experiments, in which the condition, $v \ll 1$ ($V \ll
kT/ze$), is routinely violated.

More significant modifications arise at leading order in the strongly
nonlinear regime (or at higher order in the weakly nonlinear regime). 
Ohm's law breaks down due to
concentration gradients, as the double layers absorb a significant
amount of neutral salt from the bulk. In two or more dimensions, the
dimensionless leading-order equations for $\bar{c}_0(\rb,t)$ and
$\bar{\phi}_0(\rb,t)$ in section~\ref{sec:strong} take the form, 
\begin{equation}
\frac{\partial \bar{c}_0}{\partial \bar{t}} = \nabla^2\bar{c}_0,
\ \ \ \del\cdot(\bar{c}_0\del\bar{\phi}) = 0 ,
\end{equation}
where we scale time to the bulk diffusion time. This assumes $\alpha_d
< 1$ so that no transient space charge layers form. 

The effective boundary
conditions still involve the small parameter, $\epsilon$, as in one dimension, since
the natural scale is the RC charging time, but there are some new
terms in higher dimensions:
\begin{eqnarray}
\epsilon\, \frac{\partial \tilde{q}_0}{\partial \bar{t}} & = & 
\nb\cdot(\bar{c}_0 \del \bar{\phi}_0) -   
\Du \, \del_s \cdot \tilde{J}_s   \label{eq:Js} \\
\epsilon\, \frac{\partial \tilde{w}_0}{\partial \bar{t}} & = & 
\nb\cdot \del \bar{c}_0 -   
\Du \, \del_s \cdot (\tilde{D}_s \del_s \tilde{w}_0)  \label{eq:Ds}
\end{eqnarray}
The last term in Eq.~(\ref{eq:Js}) is the surface divergence of the
(leading-order) dimensionless 
tangential current, $\tilde{J}_s$, in the diffuse layer; the size of this term compared
to the normal current is governed by a Dukhin number, based on the
largest expected total zeta potential. Similarly, the last term in
Eq.~(\ref{eq:Ds}) is the surface divergence of the tangential diffusive
flux in the diffuse layer, where $\tilde{D}_s$ is a dimensionless surface
diffusivity; again, this term is of order $\Du$ smaller than the
normal diffusive flux.  

Formulae for $\tilde{J}_s$ and $\tilde{D}_s$ can be derived
systematically using the matched asymptotic expansions, which is
beyond the scope of this paper. The classical results of
Bikerman~\cite{bikerman33,bikerman35} and Deryagin and
Dukhin~\cite{deryagin69} are available for the case of weak applied
voltages ($\zeta_{ind}\ll kT/e$) and large equilibrium surface charges
($\zeta_{eq} > kT/e$, $\Du(\zeta_{eq})\approx 1$), and many Russian
authors have studied electrokinetic phenomena in this
regime~\cite{dukhin80,dukhin93}.  The case of strongly nonlinear
dynamics ($\zeta_{ind} > kT/e$,  $\alpha_d(\zeta_{ind}) \approx 1$),
however, should be revisited in more detail to see if any changes
arise for strong, time-dependent applied voltages. We suggest as a
basic open question analyzing the electrochemical response of a metal
cylinder or sphere in a strong, suddenly applied, uniform background
DC field.

Another interesting issue is the stability of our one-dimensional
solution. One should consider small space-dependent perturbations of
the solution at various large voltages, in both the weakly nonlinear
and strongly nonlinear regimes. The general transient analysis in two
or more dimensions with the same equations and boundary conditions
presents an interesting challenge.

\subsection{ General Electrolytes and Faradaic Reactions }

Even in one dimension, it would be interesting to extend our analysis
to more general situations involving asymmetric or multicomponent
electrolytes, which undergo Faradaic processes at electrode surfaces. 
Restoring dimensions, the bulk electrolyte is described by the $N$ ionic
concentrations, $C_i$, $i=1,2,\ldots,N$, satisfying mass conservation,
\begin{equation}
\frac{\partial C_i}{\partial \tau} = - \del\cdot\Fb_i
\end{equation}
where $\Fb_i$ is the flux density due to diffusion and
electromigration,
\begin{equation}
\Fb_i = - D_i \del C_i - \mu_i z_i e C_i \del \Phi  
\end{equation}
 as in Eq.~(\ref{eq:eqdim}). For thin double layers, at leading order 
the bulk remains neutral (as long as $\alpha_d < 1$ to avoid space 
charge formation), so the potential is determined implicitly by the condition of
electroneutrality, 
\begin{equation}
\rho_e = \sum_{i=1}^N z_i e C_i = 0 .
\end{equation}
These are the standard equations of bulk
electrochemistry~\cite{newman}, but interesting physical effects are
contained in the effective boundary conditions. 

Generalizing the total surface charge density $q$ and excess surface
concentration $w$, we define $\Gamma_i$ to be the surface
concentration of species $i$ absorbed in the diffuse layer. To be precise, it
is the integral of the leading-order excess concentration relative to
the bulk over the inner coordinate, as in Eqs.~(\ref{eq:qt}) and
(\ref{eq:w}). For example, $q = ze(\Gamma_+ - \Gamma_-)/2$ and $w = (\Gamma_++
\Gamma_-)/2$ for a symmetric binary electrolyte. 

Following the procedure above, the boundary
conditions on the leading-order bulk approximation are of the form:
\begin{equation}
-\frac{\partial \Gamma_i}{\partial \tau} = \nb\cdot \Fb_i +
\del_s\cdot\Fb_{si} + R_i
\end{equation}
where $\Fb_{si}(C_i,\Phi)$ is the surface flux density of species $i$ in the
double layer~\cite{deryagin69} and $R_i(\{C_i\},\Phi)$ is the reaction-rate density
for any Faradaic processes consuming (or producing) species $i$ at the
surface. The usual assumption for $R_i$ involves Arrhenius kinetics,
as in the Butler-Volmer equation, but the Frumkin correction for
concentration variations across the diffuse layer must be taken into
account~\cite{newman,bard}. 

The general system of nonlinear equations is challenging to solve, even
numerically, due to multiple length and time scales. Boundary-layer
theory provides only a partial simplification by integrating out the
smallest length scale. As described in
section~\ref{sec:history}, various special cases of the effective
equations have been considered in the literature,
but much remains to be done, especially for strongly nonlinear
dynamics in large applied voltages. In micro-electrochemical or
biological systems, this regime is easily reached, so it merits
additional mathematical study and comparison with experimental data,
in part to test the applicability of the Nernst-Planck equations in
micro-systems. Another interesting aspect 
is the coupling of electrochemical dynamics to fluid flow, which is
finding new applications in microfluidic devices.

\acknowledgments

This work was supported in part by the MRSEC Program  of the National
Science Foundation under award number DMR 02-13282 (MZB), by ESPCI
through the Paris Sciences Chair (MZB), and by an ACI from the
Minist\`ere de la Recherche (AA). We are grateful to K. Chu and the
MIT librarians for help in obtaining articles from the ``ancient''
and Russian literatures.

%\references

\end{document}